%% file: forest2d.tex
\magnification=\magstep1
\voffset=-1.0pc
\vsize=48.5pc
\input psfig.sty
\input ffm_macros

\def\age{\,\raise2pt\hbox{$\mathop{>}\limits_{\raise 2pt
\hbox{$\sim$}}$}\,}
\def\ale{\,\raise2pt\hbox{$\mathop{<}\limits_{\raise 2pt
\hbox{$\sim$}}$}\,}
\def\sn{\smallskip\noindent}
\def\mn{\medskip\noindent}
\def\bn{\bigskip\noindent}
\def\min{{\rm min}}
\def\max{{\rm max}}
\def\abs#1{\vert #1 \vert}

\def\pr{p_{{\rm re}}}
\def\plin{p_{{\rm lin}}}
\def\xv{{\vec{x}}}
\def\yv{{\vec{y}}}
\chapternumstyle{blank}                           
\sectionnumstyle{arabic}                          
\def\cite#1{$\lbrack#1\rbrack$}
\def\bibitem#1{\parindent=9mm\item{\hbox to 7 mm{\cite{#1}\hfill}}}
\def\figindents{\leftskip=4.5 true pc \rightskip=4 true pc}
\def\BTWa{1}
\def\BTWb{2}
\def\nature{3}
\def\BakS{4}
\def\FBS{5}
\def\PMB{6}
\def\HenOld{7}
\def\DrSchwA{8}
\def\BCT{9}
\def\GrKa{10}
\def\DMS{11}
\def\Henley{12}
\def\Grassb{13}
\def\CDrSchE{14}
\def\HoPe{15}
\def\NeSn{16}
\def\BJW{17}
\def\LVZ{18}
\def\WWW{19}
\def\ClarP{20}
\def\CFO{21}
\def\CDrSchw{22}
\def\StAh{23}
\def\DrSchF{24}
\def\Drossel{25}
\def\LPS{26}
\def\MueK{27}
\def\CaNa{28}
\def\href#1#2{{#2}}
%
%
\font\large=cmbx12 scaled \magstep2
\font\bigf=cmr10 scaled \magstep2
\pageno=0
\def\folio{
\ifnum\pageno<1 \footline{\hfil} \else\number\pageno \fi}
\rightline{cond-mat/9609105}
\rightline{September 1996}
\rightline{Version of October, $23^{{\rm th}}$ 1996}
\vskip 3.0truecm
\centerline{\large Length Scales and Power Laws in}
\vskip 0.5truecm
\centerline{\large the Two-Dimensional Forest-Fire Model}
\vskip 1.5truecm
\centerline{\bigf A.\ Honecker and I.\ Peschel}
\bigskip\medskip
\centerline{\it Fachbereich Physik, Freie Universit\"at Berlin,}
\centerline{\it Arnimallee 14, D--14195 Berlin, Germany}
\vskip 1.9truecm
\centerline{\bf Abstract}
\vskip 0.2truecm
\noindent
We re-examine a two-dimensional forest-fire model via Monte-Carlo
simulations and show the existence of two length scales
with different critical exponents associated with clusters
and with the usual two-point correlation function of trees. We
check resp.\ improve previously obtained values for other critical
exponents and perform a first investigation of the critical
behaviour of the slowest relaxational mode. We also
investigate the possibility of describing the
critical point in terms of a distribution of the
global density. We find that some qualitative features such
as a temporal oscillation and a power law of the cluster-size
distribution can nicely be obtained from such a model that
discards the spatial structure.
\vskip 1.7cm
\centerline{to appear in Physica {\bf A}}
\vfill
\leftline{{\bf Keywords:} forest-fire model, critical phenomena, non-equilibrium systems}
\vskip 3mm
\leftline{\hbox to 5 true cm{\hrulefill}}
\leftline{e-mail:}
\leftline{\quad honecker@physik.fu-berlin.de}
\leftline{\quad peschel@aster.physik.fu-berlin.de}
\eject
\section{Introduction}
\mn
Dynamical systems that are naturally close to or on their critical point
have been proposed as an explanation of the appearance of power laws in
nature \cite{\BTWa,\BTWb}. This `self-organized criticality' still
consists to a large extent
in the study of toy models and has many open or controversial
questions. The lattice models can be grouped at least into three classes.
A first class contains models with a local conservation law like
the famous sandpile-model \cite{\BTWa,\BTWb}. Recent experiments
on piles of rice \cite{\nature} partially confirm the general theoretical
predictions by exhibiting power laws, but also show that the existence
of power laws in real systems depends on microscopic details such as the
aspect ratio of grains of rice. Models of evolution \cite{\BakS,\FBS}
are one example in a second class where the dynamics is specified
in terms of a globally selected extremal site. This extremal dynamics
can be used to obtain some general statements for the complete class
of models \cite{\PMB}. A model for self-organized criticality
that is known as the `forest-fire model' is a member of a third class
of models that have parameters which can be tuned close to the critical
point in a natural manner, but unlike in the two previous classes cannot
be entirely discarded (In dealing with this model one should be aware
that it is highly idealized and not expected to describe real
forest fires). The precise version that
we study in this paper has first been introduced in a short note
\cite{\HenOld} and arises as a certain limit of the more general model
proposed later independently in \cite{\DrSchwA}.
\mn
The two-dimensional forest-fire model has been already discussed
very controversially, mainly on the basis of Monte-Carlo simulations
where usually the accuracy of the predictions was the issue.
An originally proposed version \cite{\BCT} did not show the desired
critical behaviour \cite{\GrKa,\DMS}, and it was necessary to introduce
lightnings \cite{\DrSchwA}. Subsequently, Monte-Carlo simulations have
several times lead to values for the critical exponents which
had to be corrected later on \cite{\DrSchwA,\Henley,\Grassb,\CDrSchE}.
Here we add to this discussion by re-examining some of the quantities
investigated only in one of the earlier works \cite{\Henley}. We were
motivated by a study of the one-dimensional case \cite{\HoPe} where
we had discovered the existence of two different length scales.
The analogous question in two dimensions has been addressed in
\cite{\Henley}, but there it seemed that the two scales are
proportional to each other. Here, we present more accurate simulations
that demonstrate these two length scales to be different also in two
dimensions. As by-products we also check or improve estimates for
other exponents (see in particular \cite{\CDrSchE}). A final subject
is the approach to equilibrium which seems not
to have been addressed in a similar way before. In a second part,
we try to discard the spatial structure and introduce a global
model similar to the one of \cite{\HoPe} for one dimension. On the
one hand, some of the qualitative features of the stationary state
at the critical point (e.g.\ the power law of the cluster-size
distribution) can nicely be described by such a global model. On the
other hand, there are discrepancies in quantitative details and the
range of such a simplified model is very limited in comparison to
the full model. We believe that this is an important point, e.g.\ because
it has been suggested in \cite{\NeSn} that power laws in nature
might arise from a global (`coherent') driving that does not see
any spatial structure. Our findings here and in the one-dimensional
case \cite{\HoPe} demonstrate that the full model is not only different
from but also richer than the simplified model. This is analogous
to the result of \cite{\BJW} that versions of certain models of evolution
with and without spatial structure lead to different results (at least
if examined closely).
\mn
We now define the model before we proceed with a presentation of our simulation
results in the next Section. The forest-fire model is defined on a
cubic lattice in $d$ dimensions.
Any site can have two states: It can either be empty or it can be
occupied by a tree. The dynamics of the model is specified by the
following update rules (following \cite{\HenOld,\Henley,\Grassb,\CDrSchE}
\footnote{${}^{1})$}{
To be precise, the simulations in \cite{\Grassb,\CDrSchE} have
been performed according to slightly different rules, because they
aimed at investigating only quantities associated with clusters.
}):
In each Monte-Carlo step first choose an arbitrary site of the lattice.
\item{a)} If it is empty, grow a tree there with probability $p$.
\item{b)} If it is occupied by a tree, delete the entire geometric
          cluster of trees connected to it with probability $f$.
          This corresponds to a lightning stroke with subsequent
          spreading of the fire.
\par\noindent
A rescaling of the probabilities $p$ and $f$ just amounts
to a rescaling of the time scale, and in particular leaves the
stationary state invariant. We exploit this to set $p=1$.
There is a critical point at $f/p = 0$, but the parameter $f/p$ is
relevant and it is not legitimate to consider the forest-fire model
precisely at this critical point (compare \cite{\LVZ}).
\bn
\section{Simulation results in two dimensions}
\mn
In the following we consider the two-dimensional version
of the forest-fire model on a quadratic lattice with
periodic boundary conditions. The linear size of the lattice
will be denoted by $L$. So the volume $V$ is given by $V = L^2$.
We use a `global' time scale, a unit of which is defined by
the number of Monte-Carlo steps needed in order to visit each
site on average once, i.e.\ a unit of global time consists
of $L^2$ Monte-Carlo steps.
\mn
In order to do the simulation efficiently also for large systems
and small $f/p$ one has to be careful and use e.g.\
bitmapping technologies. For details on the implementation
compare the WWW page \cite{\WWW}. Using this program,
the simulations of this paper took about four months of CPU time on 
150MHz DEC alpha workstations.
\mn
We have investigated mainly systems of linear size $L=16384$
and parameter values $10^{-2} \ge f/p \ge 10^{-4}$.
For a simulation, one random initial condition with density
$\rho = 1 / 2$ was chosen. In order to equilibrate
the system, it was left to evolving freely for at least 15 global
time units. This equilibration time was increased to 25 global
time units for $f/p \le 3 \cdot 10^{-4}$ and to 35 global time units
for $f/p = 1 \cdot 10^{-4}$ (these times were adjusted according
to the observed time evolution of $\rho$, see also Section 2.3
below). After this, the system was iterated
further for another 60 to 90 global time units (120 units for
$f/p = 3 \cdot 10^{-4}$). During this period, measurements
were made as global averages at intervals of usually one
global time unit (for the density 200 times more frequently).
This amounts to at least $60 L^2$ measurements for each quantity
of interest.  Note that we use only a single run.
\bn
\subsection{Correlation functions}
\mn
Let us first discuss the simulation results for the usual
tree correlation function $\langle T(\xv) T(\xv+\yv)\rangle$
($T(\xv) = 1$ in a configuration with a tree at site $\xv$,
$T(\xv) = 0$ if the site $\xv$ is empty).
We have only investigated displacements $\yv$ along the vertical
axis (which can be treated particularly efficiently using bitmaps).
The treatment of the data resulting from the simulation is
straightforward because it can nicely be fitted by
$$C(y) :=
\langle T(\xv) T(\xv+y \vec{e}_2)\rangle - \langle T(\xv) \rangle^2 = a e^{-{y / \xi}}
  \label{defTp}$$
in a suitable interval $y_\min \le y \le y_\max$.
The parameters $a$ and $\xi$ were estimated by taking the logarithm of
the r.h.s.\ of \ref{defTp} and then performing a linear regression for
$y_\min \le y \le y_{\rm max}$. These bounds are chosen
such that the approximation \ref{defTp} by a single exponential function
is good. The lower cutoff can be chosen small ($y_\min \approx 20$)
independent of $f/p$. An upper cutoff $y_{\rm max}$ has to be imposed
at distances of 4 to 7 times the correlation length $\xi$ because then
statistical errors become large.
\mn
\centerline{\psfig{figure=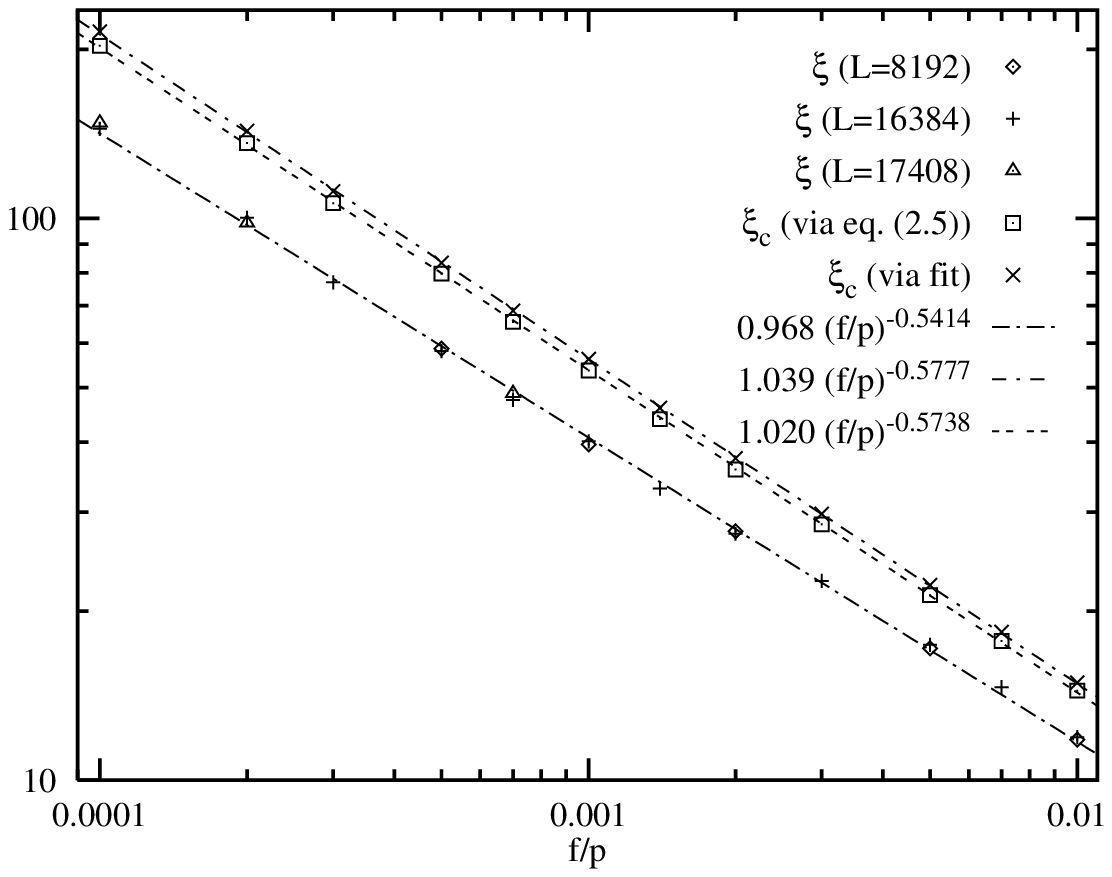}}
\sn
{\par\noindent\figindents
{\bf Fig.\ 1:}
The two correlation lengths $\xi$ and $\xi_c$.
For $\xi_c$ all estimates are on lattices with $L=16384$ and
the different symbols correspond to different ways of treating the
data.
\par\noindent}
\mn
Fig.\ 1 shows the results for $\xi$ obtained in this manner
on lattices with $L=8192$ (diamonds), $L=16384$ (crosses)
and $L=17408$ (triangles).
One observes that they are in good agreement with the form
$$\xi \sim \left({f \over p}\right)^{-\nu_T} \, .
  \label{DEFnuT}$$
Performing linear regression fits on a doubly logarithmic scale
one finds \footnote{${}^{2})$}{
Error estimates are always the $1 \sigma$ confidence interval of
a fit unless discussed explicitly.}
$$\nu_T = 0.541 \pm 0.004\, .
  \label{RESnuT}$$
This is close to the result $\nu_T = 0.56$ found in \cite{\Henley}.
Comparing the values for $\xi$ obtained from simulations
on different lattice sizes one can see that they may have a residual
statistical error of around 2\% which is about the same as the
scattering of the individual data points around the line
\ref{DEFnuT}. Since also further
data for larger $f/p$ is consistent with the form \ref{DEFnuT}
and the value \ref{RESnuT}, this result for $\nu_T$ and its error
bound can be considered reliable.
\mn
The estimates for the normalization constant $a$ in \ref{defTp} are
compatible with a value $a = 0.030 \pm 0.001$ independent of
$f/p$. So, in the limit $f/p \to 0$ the two-point function seems
to tend to $\langle T(\xv) \rangle^2 + a$ for $y \age 100$. In terms
of the alternative ansatz $C(y) = a y^{-\eta_{{\rm occ}}} e^{-{y / \xi}}$
used in \cite{\Henley} this corresponds to $\eta_{{\rm occ}} = 0$,
while the result found there was $\eta_{{\rm occ}} = 0.120 \pm 0.015$.
We have also looked into the possibility of a power-law correction factor
using our data for $L=16384$ and $f/p \le 3 \cdot 10^{-4}$. With the
estimates for $\xi$ shown in Fig.\ 1 one finds that
$e^{{y / \xi}} C(y) \sim y^{-0.11}$ for $y \le 20$, i.e.\ for small $y$
there is indeed a power-law correction factor
with an exponent that is consistent with \cite{\Henley}. On the other
hand, for $50 \le y \le 4 \xi$, the function $e^{{y / \xi}} C(y)$
is flat -- its smallest values are around $0.027$ and its maximal
values around $0.031$. This clearly contradicts a power-law correction
factor with $\eta_{{\rm occ}} \approx 0.11$ in the large-distance
asymptotics. Thus, for the large-distance behaviour
$\eta_{{\rm occ}}  = 0$ seems to be correct, while the power-law
correction factor observed in \cite{\Henley} applies to small distances.
\mn
We now look at a second quantity, namely
the `connected correlation function' $\langle T(\xv) T(\xv+\yv)\rangle_c$
describing the probability to find two trees at positions
$\xv$ and $\xv+\yv$ {\it inside the same cluster}. This
quantity is usually referred to as {\it the} two-point function
in the context of self-organized criticality and often is also the
only correlation function that is investigated. We demonstrated in one
spatial dimension \cite{\HoPe} that the length scales associated
to this correlation function and $C(y)$ have different critical exponents
while \cite{\Henley} suggested that in two dimensions length scales
associated to different quantities are equivalent. This latter
suggestion was based on simulations with lattice sizes up to
$512 \times 512$ and $f/p \age 5 \cdot 10^{-4}$.
One of the main aims of the simulations presented here is to
check if different length scales can be exhibited also in
two dimensions after sufficiently improving the accuracy.
As before, we restrict to displacements $\yv$ along the vertical
axis, i.e.\ we consider
$$K(y) := \langle T(\xv) T(\xv+y \vec{e}_2)\rangle_c \, . 
  \label{DEFconTPF}$$
This quantity can be determined in the same run as
the two-point function $\langle T(\xv) T(\xv+\yv)\rangle$,
but it involves the additional effort of determining all
clusters present in the system.
\mn
Following \cite{\CDrSchE} we associate a correlation length $\xi_c$
to the second moment of $K(y)$ via
$$\xi_c^2 := {\sum_{y=1}^{\infty} y^2 K(y) \over
              \sum_{y=1}^{\infty} K(y)} \, .
  \label{DEFxiC}$$
The squares in Fig.\ 1 show values of $\xi_c$ extracted from
simulations with $L=16384$ using this definition. One sees that these
values are consistent with
$$\xi_c \sim \left({f \over p}\right)^{-\nu} \, .
    \label{DEFnuC}$$
and a (preliminary) value of
$$\nu = 0.5738 \pm 0.0013 \, .
    \label{RESnuCpre}$$
This agrees roughly with the value $\nu = 0.60$ found in \cite{\Henley},
and also with the more accurate result in \cite{\CDrSchE}
which, including error bounds, is given by $\nu = 0.580 \pm 0.003$
\cite{\ClarP}. It should be noted that the lower bound $y=1$
in \ref{DEFxiC} is crucial. Starting the summation instead e.g.\
at $y=10$, one finds $\xi_c \approx 24$ at $f/p = 10^{-2}$
and $\xi_c \approx 219$ for $f/p = 10^{-4}$ instead of
$\xi_c \approx 14$ and $\xi_c \approx 203$, respectively. This in turn
would make the value for $\nu$ smaller. We will argue soon that
the start of summation $y=1$ is indeed the correct choice, but
the impact of a modification here could give rise to a slightly
larger error than the $1\sigma$ interval for the fit given in
\ref{RESnuCpre}.
\mn
We now proceed with a more detailed discussion of the
form of $K(y)$ which will also justify the definition
\ref{DEFxiC}. Eq.\ \ref{DEFxiC} is based on the following expected
form of $K(y)$:
$$K(y) = a_c \; y^{-\eta} \; e^{-y/\xi_c} \, .
\label{DEFeta}$$
In particular, if $e^{y/\xi_c} \; K(y)$ agrees well with
a power law, the determination of $\xi_c$ via \ref{DEFxiC} is justified.
Using the values of $\xi_c$ given by the boxes in Fig.\ 1 one finds that 
$e^{y/\xi_c} \; K(y)$ does indeed agree well with
$a_c \; y^{-\eta}$ for $y$ between 1 and several times $\xi_c$
\footnote{${}^{3})$}{
The values of $a_c$ are all compatible with the $f/p$-independent
value $a_c = 0.20 \pm 0.01$. Therefore, at the critical point the
power law $K(y) = a_c y^{-\eta}$ is expected to
be valid for all $y$.}.
The crosses in Fig.\ 2 show these estimates for $\eta$. 
They obviously depend on $f/p$ contrary to what one would
naively expect. This points to systematic errors in the
determination of $\eta$ which one may expect to become less
important for smaller $f/p$ where the power law is better
visible. This suggests to make a `scaling ansatz'
$\eta(f/p) = \eta + \bar{a} \; (f/p)^{\bar{b}}$
to determine the value of $\eta$ in the limit $f/p \to 0$.
A least-squares fit gives the dotted line in Fig.\ 2.
One finds a critical $\eta = 0.374 \pm 0.011$.
\mn
\centerline{\psfig{figure=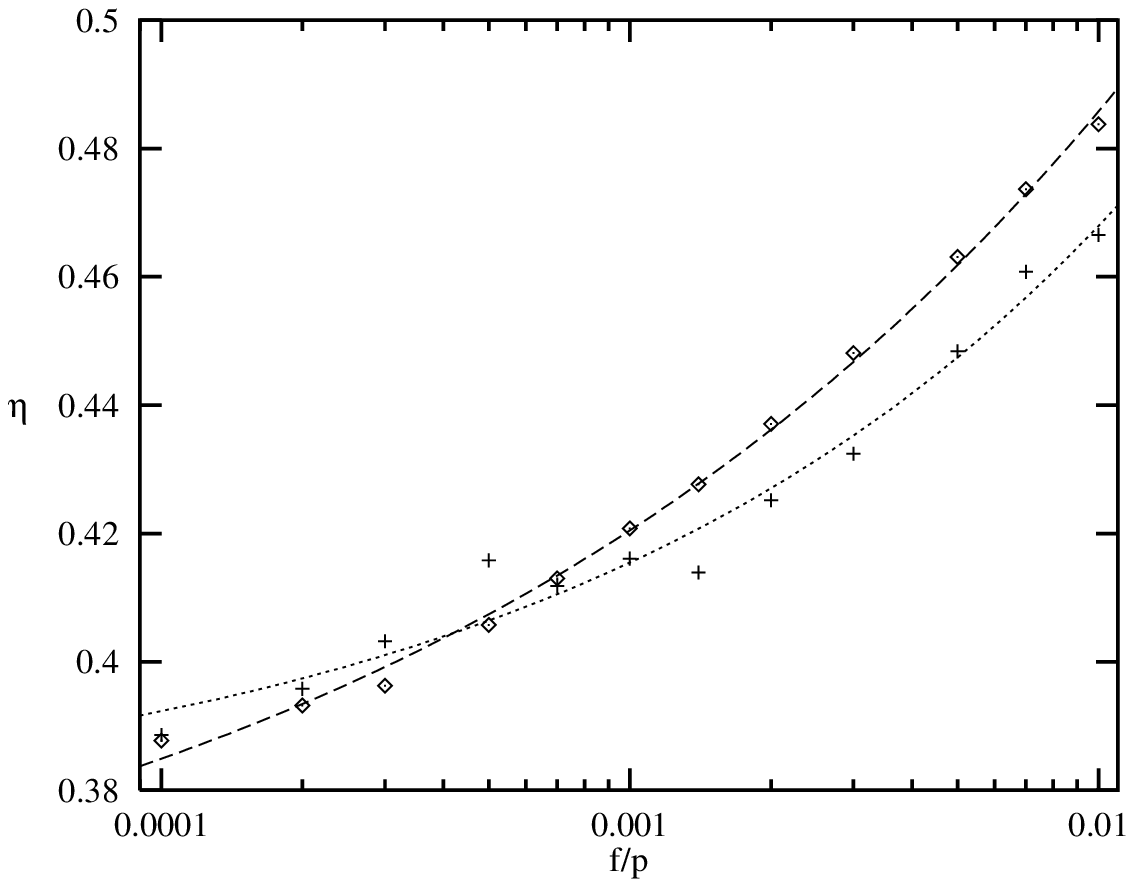}}
\sn
{\par\noindent\figindents
{\bf Fig.\ 2:}
Values for the exponent $\eta$ obtained in two different
ways (crosses and diamonds) and scaling fits (lines).
\par\noindent}
\mn
Alternatively, one can fit all three parameters in \ref{DEFeta}
directly from the data. The values for $\eta$ obtained
with this approach are shown in Fig.\ 2 by diamonds.
A scaling analysis (the line with long dashes in Fig.\ 2) yields a
critical $\eta = 0.342 \pm 0.008$. Comparing this with the value
found earlier, one finds a small disagreement. So, the error
bounds of these two estimates for the critical $\eta$ may be a
little too optimistic because they do not include
systematic errors. In order to be careful we give a final result
that includes both estimates for $\eta$ and the two error ranges:
\goodbreak
$$\eta = 0.36 \pm 0.03 \, .
\label{RESeta}$$
This is compatible with the result $\eta = 0.411 \pm 0.02$ of
\cite{\Henley} (although the error estimate of \cite{\Henley}
seems a little too optimistic, possibly because it does not
take systematic errors into account).
One also obtains different estimates for $\xi_c$ from the same
fits that gave the alternative values for $\eta$. These values
are shown in Fig.\ 1 by `$\times$'. One sees that they
are again compatible with the form \ref{DEFnuC}, and obtains
an alternative estimate for the associated critical exponent:
$\nu = 0.5777 \pm 0.0013$. There is a minor difference between
this value and the earlier estimate \ref{RESnuCpre} indicating
that we have indeed neglected systematic errors. Therefore we
give a final result 
$$\nu = 0.576 \pm 0.003
    \label{RESnuC}$$
which includes the two direct estimates and their error bound. This
final result is in excellent agreement with \cite{\CDrSchE}.
\mn
Comparing \ref{RESnuC} with \ref{RESnuT}, we see that
$\nu \ne \nu_T$, so that $\xi_c$ and $\xi$ are basically
different lengths. A difference $\nu - \nu_T \approx 0.04$ was
already observed in \cite{\Henley}, but attributed to numerical
errors because it was unexpected.
\mn
It should be mentioned that our results \ref{RESeta} and \ref{RESnuC}
for $\eta$ and $\nu$ do not satisfy the scaling relation $2-\eta = 1/\nu$
\cite{\Henley}. We can only speculate about the reason for the
disagreement. For example, in the derivation of this scaling
relation one assumes that $\eta$ and $\nu$ do not depend on
the direction of the displacement vector $\yv$, and we have
not checked whether this is indeed true. Another possibility
is that we have still overlooked systematic errors. Inserting
$\nu$ (which is the more reliable value) according to \ref{RESnuC}
into the scaling relation $2-\eta = 1/\nu$ yields $\eta \approx 0.28$
which is possible if our error estimate in \ref{RESeta}
is by a factor of about 3 too small.
\bn
\subsection{Cluster-size distribution}
\mn
The distribution $n(s)$ of clusters with size $s$ arises as
a by-product of the determination of $K(y)$ during the simulations.
We extract an exponent $\tau$ from it following the lines
described in detail in \cite{\Henley,\Grassb}. One introduces
the quantity $P(s) = \sum_{s' > s} s' n(s')$. Assuming that
it behaves as $P(s) = \alpha s^{2-\tau} e^{-s/s_{{\rm max}}}$,
one can use three-parameter fits to obtain estimates for $\tau$
and $s_{{\rm max}}$. Extrapolation of the values obtained in
this manner from the simulations with $L=16384$ yields
$\tau = 2.1595 \pm 0.0045$ for $f/p \to 0$.
The values of $\tau$ for $f/p > 0$ approach
this limiting value from above. An alternative way to extract
a value of $\tau$ is to look directly at $n(s)$ and assume
that $n(s) \sim s^{- \tau}$ for intermediate $s$. Applying this
second method to the same data and to some results
for $L=8192$ we obtain the estimate $\tau = 2.159 \pm 0.006$
at the critical point. This limit is now approached from below and
is in excellent agreement with the one obtained
before. To be on the safe side we retain the value with the
larger error bound as the final result:
$$\tau = 2.159 \pm 0.006 \, .
\label{REStau}$$
This value agrees within error bounds with most previous results
\cite{\Henley,\CFO,\Grassb,\CDrSchE}, but our error bound is
considerably smaller than in most of them.
\mn
When one extracts estimates for $\tau$ from $P(s)$ one also
obtains estimates for $s_{{\rm max}}$. These values are in
good agreement with the form $s_{{\rm max}} \sim (f/p)^{-\lambda}$
with $\lambda = 1.171 \pm 0.004$. To check
the validity of our error estimate we can use the scaling
relation $\lambda = 1/(3-\tau)$ \cite{\CDrSchE}. After inserting
\ref{REStau} one finds the prediction $\lambda = 1.189 \pm 0.008$.
This prediction agrees roughly with the direct estimate. However,
there is a small discrepancy indicating that the estimate
$$\lambda = 1.17 \pm 0.02
\label{RESlam}$$
is more realistic. Within error bounds we find good agreement
with the values of \cite{\CDrSchE} and \cite{\Henley}. From
the results \ref{RESnuC} and \ref{RESlam} we
obtain the fractal dimension $\mu = 2.03 \pm 0.04$ using
the scaling relation $\mu = \lambda/\nu$ \cite{\CDrSchE}. This
does not quite agree with \cite{\CDrSchE} where $\mu = 1.96 \pm 0.01$
was found, but would favour instead the expectation $\mu = 2$
of \cite{\Grassb}. Unfortunately, with our simulations we had not
aimed at determining $\mu$ and we are therefore not able to clarify
this interesting point.
\bn
\subsection{Density and time evolution}
\mn
Here we study the critical behaviour of the stationary density
as well as the temporal behaviour of the density. The latter yields
the lowest gap in the spectrum of the time-evolution operator which is
given by a straightforward generalization of eq.\ (3.6) in
\cite{\HoPe}. Some high relaxational modes of this time-evolution
operator can be written down explicitly as we demonstrate in Appendix A. 
In one dimension, the low-lying spectrum of this operator could be
studied numerically \cite{\HoPe}, but this is not feasible in
higher dimensions. Fortunately, simulations of $\rho(t)$ exhibit
much clearer features in two dimensions than was the case in
one dimension. This is illustrated by Fig.\ 3 which shows the
initial time evolution of $\rho(t)$ after the system was started
at $\rho(0) = 1/2$.
\mn
\centerline{\psfig{figure=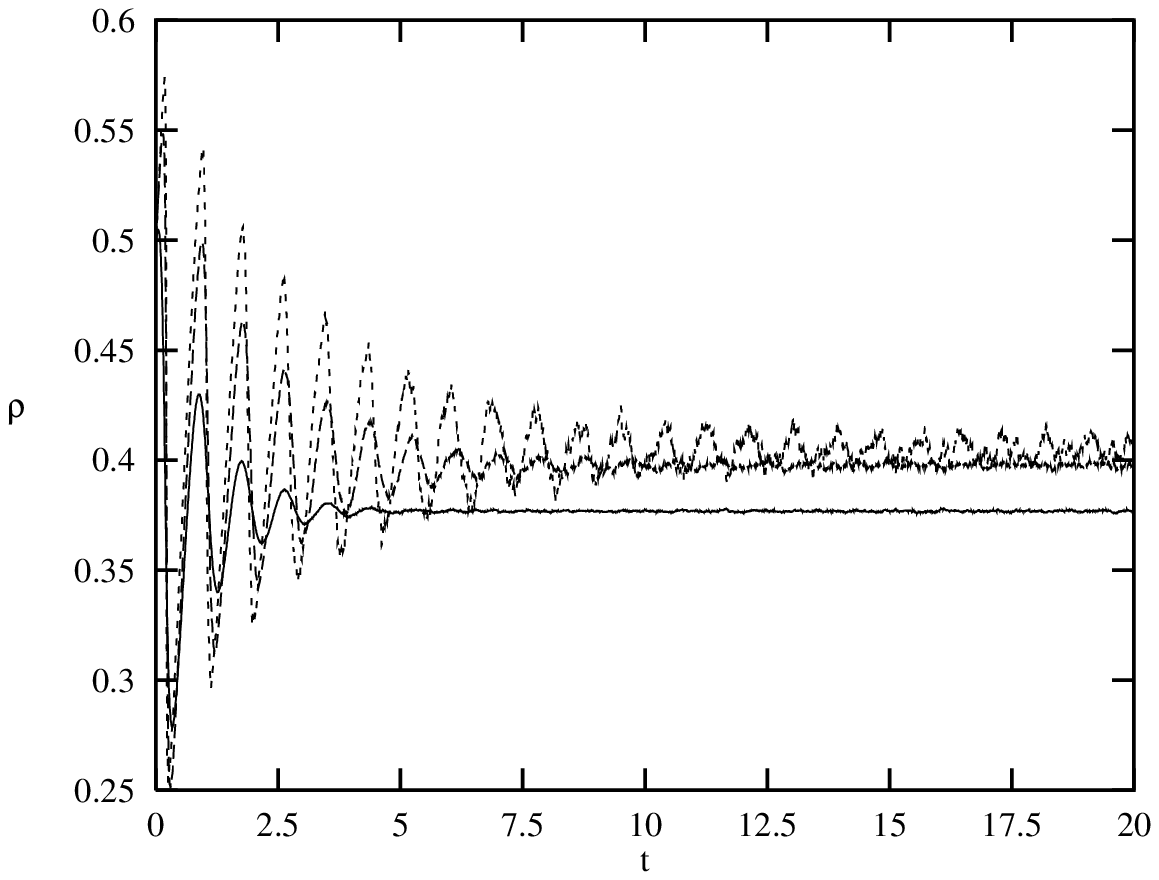}}
\sn
{\par\noindent\figindents
{\bf Fig.\ 3:}
The density $\rho(t)$ as a function of global time $t$. Shown are
results from simulations with $f/p = 10^{-2}$ (bottom, full line),
$f/p = 10^{-3}$ (line with long dashes, middle) and $f/p = 10^{-4}$
(top line with short dashes).
\par\noindent}
\mn
First we examine the critical behaviour of the stationary density
of trees $\rho(\infty)$. The values in Table 1 are obtained
by averaging $\rho(t)$ over times $t$ after the equilibration time,
i.e.\ at or beyond the right border of Fig.\ 3.
\mn
\centerline{\vbox{
\hbox{
\vrule \hskip 1pt
\vbox{ \offinterlineskip
\def\tablespace{height2pt&\omit&&\omit&&\omit&&\omit&\cr}
\def\tablerule{ \tablespace
                \noalign{\hrule}
                \tablespace        }
\hrule
\halign{&\vrule#&
  \strut \quad\hfil#\hfil \quad\cr
\tablespace
\tablespace
& $f/p$    && $\rho(\infty)$ && oscillation period $T_{{\rm osc}}$ && decay time $T$
     & \cr \tablespace \tablerule
& $1 \cdot 10^{-2}$ && 0.376833 && $0.878 \pm 0.028$ && $0.988$ & \cr \tablespace
& $7 \cdot 10^{-3}$ && 0.381539 && $0.889 \pm 0.043$ && $1.155$ & \cr \tablespace
& $5 \cdot 10^{-3}$ && 0.385406 && $0.890 \pm 0.014$ && $1.312$ & \cr \tablespace
& $3 \cdot 10^{-3}$ && 0.390296 && $0.878 \pm 0.012$ && $1.588$ & \cr \tablespace
& $2 \cdot 10^{-3}$ && 0.393463 && $0.883 \pm 0.017$ && $1.719$ & \cr \tablespace
& $1.4\cdot10^{-3}$ && 0.395793 && $0.883 \pm 0.025$ && $1.966$ & \cr \tablespace
& $1 \cdot 10^{-3}$ && 0.397672 && $0.879 \pm 0.026$ && $2.296$ & \cr \tablespace
& $7 \cdot 10^{-4}$ && 0.399321 && $0.878 \pm 0.011$ && $2.455$ & \cr \tablespace
& $5 \cdot 10^{-4}$ && 0.400673 && $0.881 \pm 0.048$ && $2.603$ & \cr \tablespace
& $3 \cdot 10^{-4}$ && 0.402291 && $0.875 \pm 0.015$ && $3.645$ & \cr \tablespace
& $2 \cdot 10^{-4}$ && 0.403231 && $0.876 \pm 0.026$ && $3.559$ & \cr \tablespace
& $1 \cdot 10^{-4}$ && 0.404696 && $0.868 \pm 0.031$ && $3.951$ & \cr \tablespace
}
\hrule}\hskip 1pt \vrule}
\hbox{\quad \hbox{Table 1:} Estimates with $L=16384$ for the stationary density of trees $\rho(\infty)$ and}
\hbox{\quad \phantom{Table 1:} the lowest decay mode in $\rho(t)$ of the two-dimensional forest-fire model.}}
}
\mn
These values are in good agreement with
$$\rho_c - \rho(\infty) \sim \left({f \over p}\right)^{1/\delta} \, ,
\label{FORMrho}$$
where the critical density $\rho_c$ and the critical exponent
$\delta$ are given by
$$\eqalignno{
\rho_c &= 0.40844 \pm 0.00011 \, , &\eqnlabel{RESrhoC} \cr
1/\delta &= 0.466 \pm 0.004 \, .  &\eqnlabel{RESdelta}
}$$
These results agree within error bounds with the
values obtained in \cite{\CDrSchE}, but our bounds are
smaller. In particular we can now rule out that
$1/\delta = 1/2$ as proposed by \cite{\Grassb}.
\mn
Having determined $\rho(\infty)$, it is straightforward to extract
the oscillation period and decay time of the slowest decay mode from
$\rho(t)$. First, one determines those $t$ where $\rho(t)$ crosses
the value $\rho(\infty)$. From the distances between these crossings
the oscillation period $T_{{\rm osc}}$ can be determined easily.
Averaging 10 to 15 half oscillation periods estimated in this
manner for a suitable interval of time in Fig.\ 3 leads to the values
given in Table 1. One observes that the oscillation period
equals $T_{{\rm osc}} = 0.88$ within error bounds for all $f/p$,
i.e.\ the slowest relaxational mode oscillates with a constant
frequency. This is to be contrasted with the one-dimensional case where
simulations of $\rho(t)$ clearly demonstrated that the oscillation
period depends on $f/p$ \cite{\HoPe}.
\mn
Finally, we extract the leading decay time $T$ from $\rho(t)$
according to the following procedure. At times $t$ precisely
in the middle between two subsequent crossings used for the
determination of the oscillation period, the value of
$\abs{\rho(t) - \rho(\infty)}$ is determined. One finds
for these (approximately ten) values of $t$ that
$\abs{\rho(t) - \rho(\infty)} \sim \exp(-t/T)$ from which
it is straightforward to obtain the estimates for $T$
presented in Table 1 \footnote{${}^{4})$}{
These estimates also verify that our
equilibration times are long enough, since we have
equilibrated the system for at least $6 T$ before starting to
collect data. So, the non-stationary modes are damped by
factors of at least $\exp(-6) \approx 2 \cdot 10^{-3}$ 
during data collection.
}. These values for $T$ are compatible
with a critical behaviour
$$T \sim \left({f \over p}\right)^{-\zeta} \, ,
\label{DEFzeta}$$
where the critical exponent $\zeta$ is determined to be
$$\zeta = 0.314 \pm 0.013 \, .
\label{RESzeta}$$
This exponent probably is another new exponent that is not
related to the ones determined so far \cite{\Henley,\CDrSchE}.
\bn
\subsection{Summary of simulations}
\mn
Table 2 summarizes our results for the critical exponents
of the two-dimensional forest-fire model. It also 
includes the critical density $\rho_c$ and the global
oscillation period which seems to be
independent of $f/p$. For comparison we have
also included the results for one dimension \cite{\HoPe,\CDrSchw}.
In that case, the oscillation period diverges and
we have listed the exponent rather than a period in Table 2.
Similarly, the amplitude $a$ vanishes in one dimension for
$f/p \to 0$ but is roughly constant in two dimensions.
\mn
\centerline{\vbox{
\hbox{
\vrule \hskip 1pt
\vbox{ \offinterlineskip
\def\tablespace{height2pt&\omit&&\omit&&\omit&\cr}
\def\tablerule{ \tablespace
                \noalign{\hrule}
                \tablespace        }
\hrule
\halign{&\vrule#&
  \strut \quad\hfil#\hfil \quad\cr
\tablespace
\tablespace
& quantity   && value in $d=2$          && value in $d=1$ & \cr \tablespace \tablerule
& $\nu_T$    && $0.541 \pm 0.004$       && $0.8336 \pm 0.0036$ & \cr \tablespace
& $\eta_{{\rm occ}}$ && $0$             && $0$ & \cr \tablespace
& $\nu$      && $0.576 \pm 0.003$       && $1$ & \cr \tablespace
& $\eta$     && $0.36 \pm 0.03$         && $0$ & \cr \tablespace
& $\tau$     && $2.159 \pm 0.006$       && $2$ & \cr \tablespace
& $\lambda$  && $1.17 \pm 0.02$         && $1$ & \cr \tablespace
& $1/\delta$ && $0.466 \pm 0.004$       && $0$ & \cr \tablespace
& $\zeta$    && $0.314 \pm 0.013$       && $\approx 0.405$ & \cr \tablerule
& $T_{{\rm osc}}$  && period: $0.88 \pm 0.02$ &&
                           exponent $\approx 0.194$ & \cr \tablespace
& $a$        && $0.030 \pm 0.001$       && $\sim (f/p)^{0.1031 \pm 0.0022}$ & \cr \tablespace
& $\rho_c$   && $0.40844 \pm 0.00011$   && $1$ & \cr \tablespace
}
\hrule}\hskip 1pt \vrule}
\hbox{\quad \hbox{Table 2:} Summary of our results for the critical behaviour of the}
\hbox{\quad \phantom{Table 2:} forest-fire model.}}
}
\vfill
\eject
\section{Global models} 
\mn
We now wish to investigate through simplified variants of
the model to what extent one can describe the
stationary properties of the two-dimensional forest-fire model
by global variables in a way similar to \cite{\HoPe}. There
it was shown that in order to describe the one-dimensional
critical stationary state it suffices to know the relative weight
of the sum of all configurations with a fixed number of occupied
(or empty) sites. Thus one is dealing with a kind of grand canonical
ideal lattice gas. Such a model has no intrinsic spatial structure
and leads to a two-point function independent of the distance.
It can therefore be used to describe the asymptotic behaviour of
the critical $C(y)$ of Section 2. Working with the (global) density
of trees $\rho$, one has to specify the probability distribution
$p(\rho)$ and obtains $C(y) = \langle \rho^2 \rangle - \langle \rho \rangle^2$
for $y \ne 0$ in the thermodynamic limit. This is positive for all
continuous distributions.
\mn
In two dimensions the limits $f/p \to 0$ and $L \to \infty$ do
not commute with each other (in contrast to one dimension).
Therefore, perturbation theory
cannot be used to compute $p(\rho)$, and in fact we do not
know of a good analytic method to determine $p(\rho)$
from the rules described in the Introduction. Therefore
we use heuristic arguments and simulations to discuss it.
\mn
The forest-fire model reminds one of site percolation. Thus,
it is natural to try to relate the forest-fire model to
percolation and gain some insight from that. Attempts in
this direction have been made e.g.\ in \cite{\DrSchwA,\CFO}. 
We will also try to use some results of percolation theory,
but we will follow a different route. Namely, we try
to interpret the stationary state of the forest-fire model at
the critical point as a suitable ensemble of percolation
problems with a distribution $p(\rho)$.
\mn
It is known from percolation theory \cite{\StAh} that in a homogeneous
configuration with a density $\rho$ above the percolation
threshold $\rho_{{\rm perc}}$ two arbitrary sites are connected
with a finite probability. In such configurations, the dynamics
of the forest-fire model with arbitrarily small $f/p > 0$
would very quickly destroy all percolating
clusters and thus drive the density below the percolation
threshold \footnote{${}^{5})$}{
This implies that the density $\rho$ cannot be continuous
at $f/p=0$ for $d>1$ and is the reason why perturbation
theory cannot be used in higher dimensions.
}. Thus, the probability $p(\rho)$ to have a global
density above the percolation threshold $\rho_{{\rm perc}}$
must vanish in the forest-fire model, i.e.\ $p(\rho) = 0$
for $\rho > \rho_{{\rm perc}}$. In one dimension one has
$\rho_{{\rm perc}} = 1$ and in two dimensions
$\rho_{{\rm perc}} = 0.592746$ \cite{\StAh}. In all simulations
$\rho(t)$ was well below this percolation threshold
at all times (compare Fig.\ 3).
\mn
It is useful to visualize the stationary state of the
forest-fire model in order to gain some intuition.
Fig.\ 4 shows an area of $760 \times 472$ sites
at $t=66$ during a simulation with $L=16384$ and $f/p = 10^{-4}$
(compare also Fig.\ 2 of \cite{\DrSchF}, Fig.\ 1 of \cite{\CDrSchE}
and Fig.\ 6 of \cite{\Grassb}). One observes that at a certain fixed time
the system consists of rather well-defined patches with different
mean density of trees. These patches are not to be confused with
single forest clusters, they usually contain many such clusters.
Their typical size increases as $f/p$ becomes
smaller, which reflects the divergence of correlation lengths.
Looking at the time evolution of such a state
\footnote{${}^{6})$}{
On X11 platforms, such a visualization is possible with the code
used for the simulations presented in this paper. This code is
available on the WWW \cite{\WWW}.},
one observes an increase of density due to growth of trees
that is constant throughout the patches and that lightning
strikes essentially only the patches with the highest density. 
After a lightning has struck such a patch, a new patch with a
low density is created. This new density is not really zero because
there are always some trees in the patch that are not connected
to the cluster which is destroyed. The idea now is
that the important information about the critical stationary state
is given by the distribution
$p(\rho)$ of densities in these patches and that nothing essential
changes if we replace such patchy systems with an ensemble of systems of
{\it global} density $\rho$ occurring with the same probability $p(\rho)$.
Of course, it remains to be tested to what extent this picture
works.
\mn
\centerline{\psfig{figure=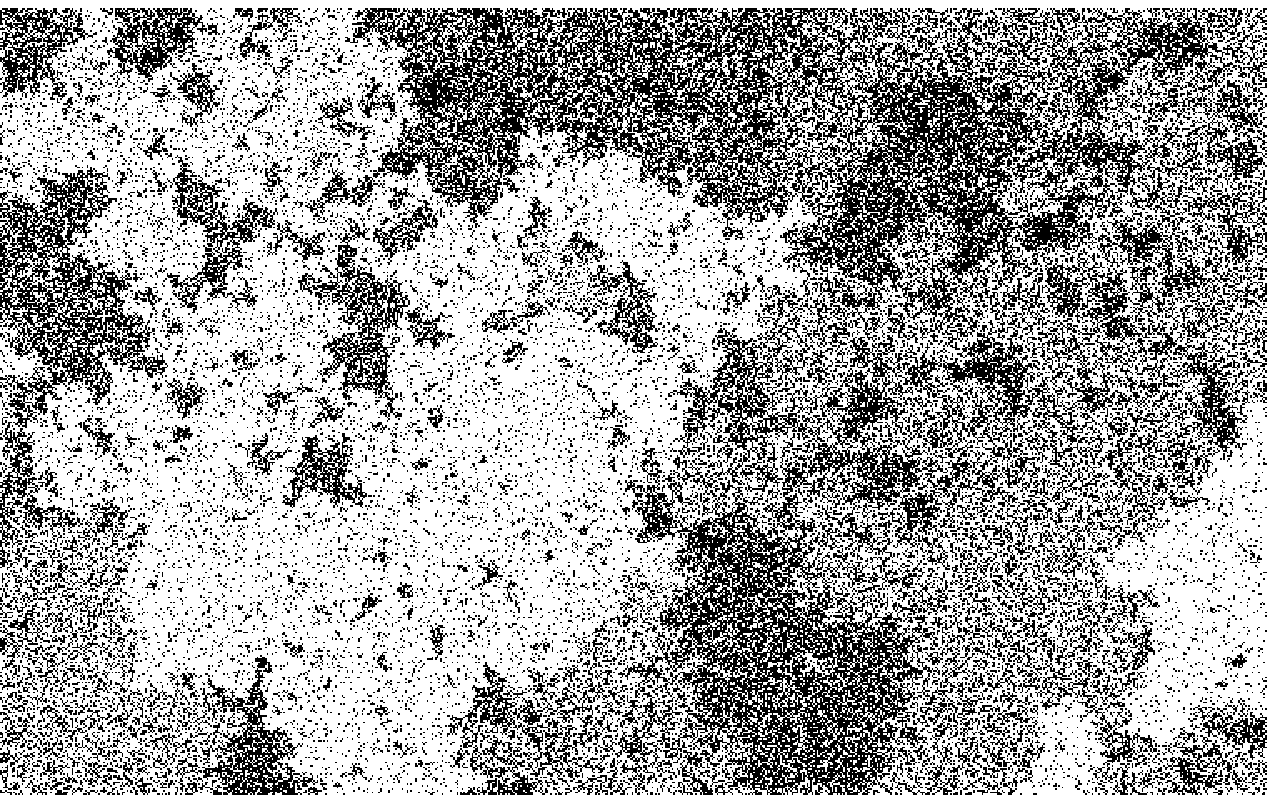}}
\sn
{\par\noindent\figindents
{\bf Fig.\ 4:}
Snapshot of an area with $760 \times 472$ sites in the stationary
state during a simulation with $L=16384$ and $f/p = 10^{-4}$.
Trees are black and empty places white.
\par\noindent}
\bn
\subsection{A simple model}
\mn
Let us make a very simple model based on these ideas.
Assume that below the percolation threshold $\rho_{{\rm perc}}$ 
trees just grow (with probability $p=1$) and no lightning strikes.
As soon as the percolation threshold is exceeded, lightning
strikes immediately and destroys all trees in the system
(not just those in the percolating cluster). In
order to determine $p(\rho)$ we first compute the mean lifetime
of a configuration with density $\rho$, assuming that no lightning
strikes (this consideration will also be useful later on). Such a
configuration lives precisely $n$ local updates if $n-1$
times an occupied place and then an empty place are selected.
Because of the above assumptions there is no correlation and
therefore the probability of this to happen is given by
$\rho^{n-1} (1-\rho)$. This yields the expectation value $t(\rho)$
of the lifetime as
$$t(\rho) = \sum_{n=1}^{\infty} n \rho^{n-1} (1-\rho) 
= {1 \over 1-\rho} \, .
\label{timeDens}$$
If no lightning strikes, the probability $p(\rho)$ to find
a configuration with density $\rho$ is proportional to the
lifetime of a state with this density. Taking into account that lightning
strikes all trees at $\rho_{{\rm perc}}$, we find for this simple model
$$p(\rho) =
\cases{{\cal N} (1-\rho)^{-1}\, , & $\rho < \rho_{{\rm perc}}$, \cr
              0              \, , & $\rho > \rho_{{\rm perc}}$, \cr}
\label{simpleGmodel}$$
with the normalization constant given by
${\cal N}^{-1} = \int_0^{\rho_{{\rm perc}}} {\rm d}\rho \; (1-\rho)^{-1}$.
In one dimension, \ref{simpleGmodel} agrees with eq.\ (5.4) of \cite{\HoPe}
which was derived there using a different argument.
\mn
This simple model has the following periodic time evolution: Trees grow
until the density reaches the percolation threshold. Then lightning
strikes and the process is restarted with a completely
empty system. The average time needed for such a cycle is precisely
the global oscillation time and is given by
$\int_0^{\rho_{{\rm perc}}} {\rm d}\rho \; t(\rho)$. Inserting
$t(\rho)$ according to \ref{timeDens} and the value of $\rho_{{\rm perc}}$
in two dimensions yields an oscillation period of
$T_{{\rm osc}} \approx 0.898$ which is in very good agreement with what
we found in simulations (compare Table 1).
The mean density is given by $\int_0^1 {\rm d}\rho \; \rho p(\rho)$
from which one finds a critical density $\rho_c = 0.340\ldots$
(this deviates notably from the result \ref{RESrhoC} found by simulations
of the full model). Finally, the probability to find two trees
at arbitrary places is given by the second moment of $p(\rho)$,
i.e.\ by $\int_0^1 {\rm d}\rho \; \rho^2 p(\rho)$. So,
the two-point function exceeds the value
$\rho_c^2$ by an amount $a=0.0289\ldots$ which agrees within
error bounds with what we found by simulations for the large-distance
asymptotics of the two-point function. Cluster-type quantities
are not accessible as easily and would e.g.\ require again
Monte-Carlo simulations.
\mn
Although this simple model yields very good values for two
quantities, its failure to give the correct $\rho_c$ is not
surprising. Firstly, we have neglected the fact that lightning
can already strike configurations with $\rho < \rho_{{\rm perc}}$
even for arbitrarily small $f/p > 0$ (compare Appendix A).
Secondly, lightning does not really lead to the completely
empty system, but usually leaves some isolated trees or
small clusters in the patch behind. Unfortunately we do not
know how to treat either effect analytically.
This lack of knowledge also has the effect that we can
in general not compute an oscillation time from $p(\rho)$
although we do of course still think of the system as
evolving in cycles (compare also section 6.3.2 of \cite{\Drossel}).
\bn
\subsection{Realistic distributions}
\mn
Next we try to obtain a realistic $p(\rho)$ from
Monte-Carlo simulations.
Measurements of the global density cannot be used to extract
$p(\rho)$ from a simulation, because $\rho$ fluctuates only very little
around its mean value for sufficiently large systems (see Fig.\ 3).
Therefore, one has to look at the distribution of local densities.
In a large system, areas with different local densities coexist
and we are interested precisely in these local fluctuations and
not just the global average.
We have decided to divide the system into $16 \times 16$
plaquettes and use the distribution of the average density
per plaquette. This size of the plaquettes was chosen because
then their linear extent is much smaller than the correlation
lengths and they contain sufficiently many sites to obtain a
fairly smooth distribution. Fig.\ 5 shows a result $\pr(\rho)$
obtained in this manner using the parameter values
closest to the critical point, namely $f/p = 10^{-4}$
and $L=16384$. Samples were taken at the same
90 times where also the correlation functions were determined,
amounting to a total of almost $10^{8}$ samples for local
densities. The normalization in Fig.\ 5 is such
that $\sum_{r=0}^{256} \pr(r/256) = 1$.
\mn
As explained above,
the first and second moments of $p(\rho)$ are related to the
mean density and the asymptotic value of the two-point function.
From this one finds
$$\rho(\infty) = 0.4044\ldots \, , \qquad a = 0.031\ldots \, .
\label{ParamRealModel}$$
The way we have determined $\pr(\rho)$ ensures that the first moment
indeed equals the value obtained by directly taking a global
average. The slight difference between \ref{ParamRealModel} and the value
for $\rho(\infty)$ in Table 1 is due to the fact that the latter is based
on a much larger amount of configurations.
As for the simple model presented before,
the prediction for $C(y) = a \approx 0.031$ agrees
within error bounds with what we expect for the asymptotics
of the two-point function at the critical point.
\mn
\centerline{\psfig{figure=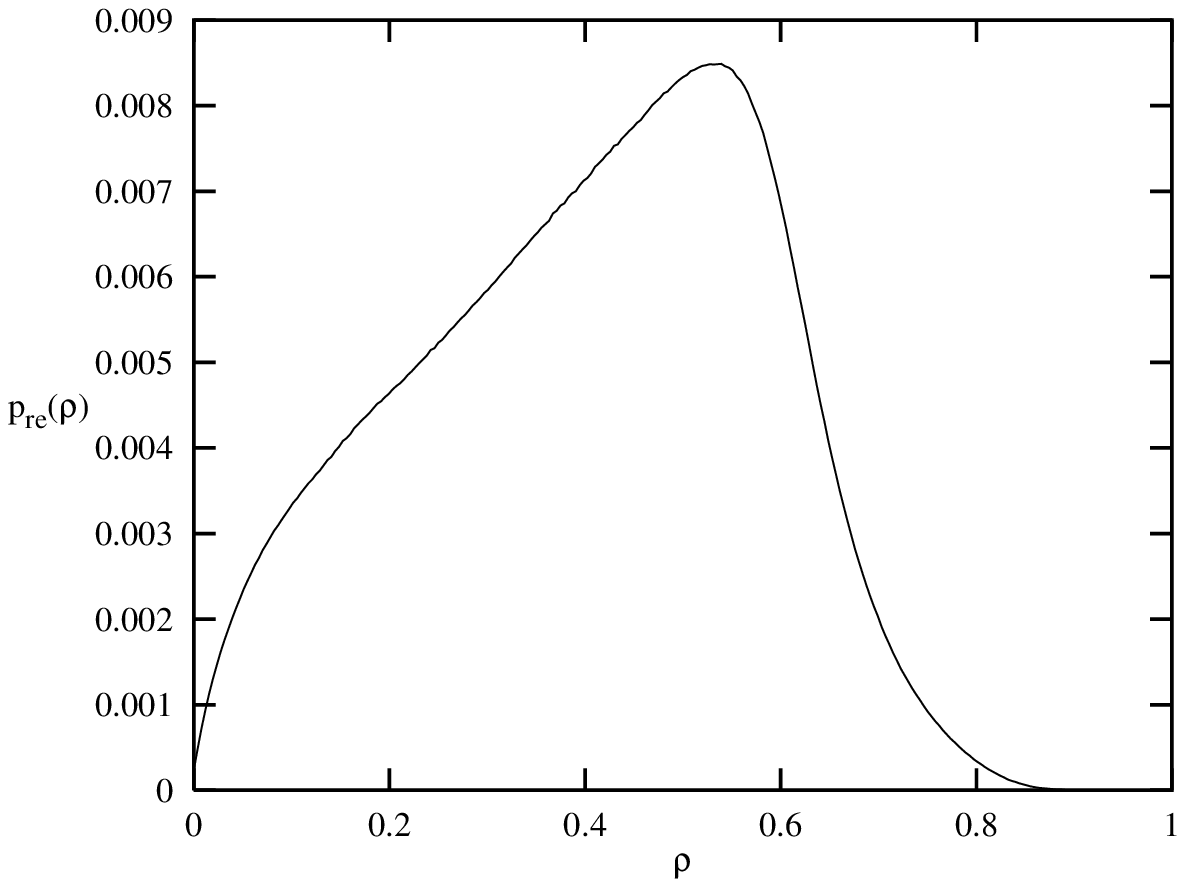}}
\sn
{\par\noindent\figindents
{\bf Fig.\ 5:}
Result for the distribution $\pr(\rho)$ on $16 \times 16$
plaquettes in a simulation with $L=16384$ and $f/p = 10^{-4}$.
\par\noindent}
\mn
Even though this basic test yields good values, one should
be aware that examining the system only in windows blurs the
distribution in Fig.\ 5 in several ways. Firstly, looking through
a window containing only 256 sites, one finds
a smearing of the density by $\Delta \rho \approx 0.02$
just because of statistical effects. Secondly, such windows may
accidentally intersect the boundary between two patches with low
and high densities. This has the effect that $p(\rho)$ is
estimated too large for intermediate $\rho$ and too small for
the extremal ones. Our choice of $16 \times 16$ plaquettes is
designed to make both effects reasonably small and is about
the best we can do.
\mn
Let us now discuss Fig.\ 5 keeping these effects in mind. 
For not too small values of $\rho$, $\pr(\rho)$ increases slowly and
reaches a maximum around $\rho \approx 0.54$. Around
$\rho \approx 0.62$ there is a sharp decrease. The broad
distribution shows that fluctuations of $\rho$ are important.
A peak just below the percolation threshold $\rho_{{\rm perc}}$
and a steep decrease above it correspond to our expectation.
However, there is still a substantial contribution to
$\pr(\rho)$ above $\rho_{{\rm perc}}$ which is not explained
by windowing effects. This is due to the patchy structure of
the system: Finite patches with $\rho > \rho_{{\rm perc}}$
are not destroyed instantly, but rather live for a time
which is the longer the smaller these patches actually are.
One could also say that the patchy structure demonstrates that
the system is actually correlated. In particular at small
distances ($y \ale 20$), the two-point function $C(y)$ retains
a $y$-depence (compare Section 2.1) for $f/p \to 0$ and
thereby exceeds the asymptotic constant $a$.
This correlation is necessary
to observe a non-trivial distribution of local densities,
but also leads to contributions to $p(\rho)$ above the
percolation threshold.
\mn
So, if we want to work with a `realistic' $p(\rho)$, the best we
can do is to proceed with the one shown in Fig.\ 5. 
Alternatively, one can work with an approximation to this
distribution where one suppresses the undesired $p(\rho)$
above the percolation threshold by hand. One such approximation
we have examined is a linear one, i.e\ $\plin(\rho) \sim \rho$ 
for $\rho < 0.59$ and $\plin(\rho) = 0$ for $\rho > 0.59$.
This linear distribution yields $\rho_c \approx 0.395$ and
$a \approx 0.019$ (both somewhat too small).
\mn
\centerline{\psfig{figure=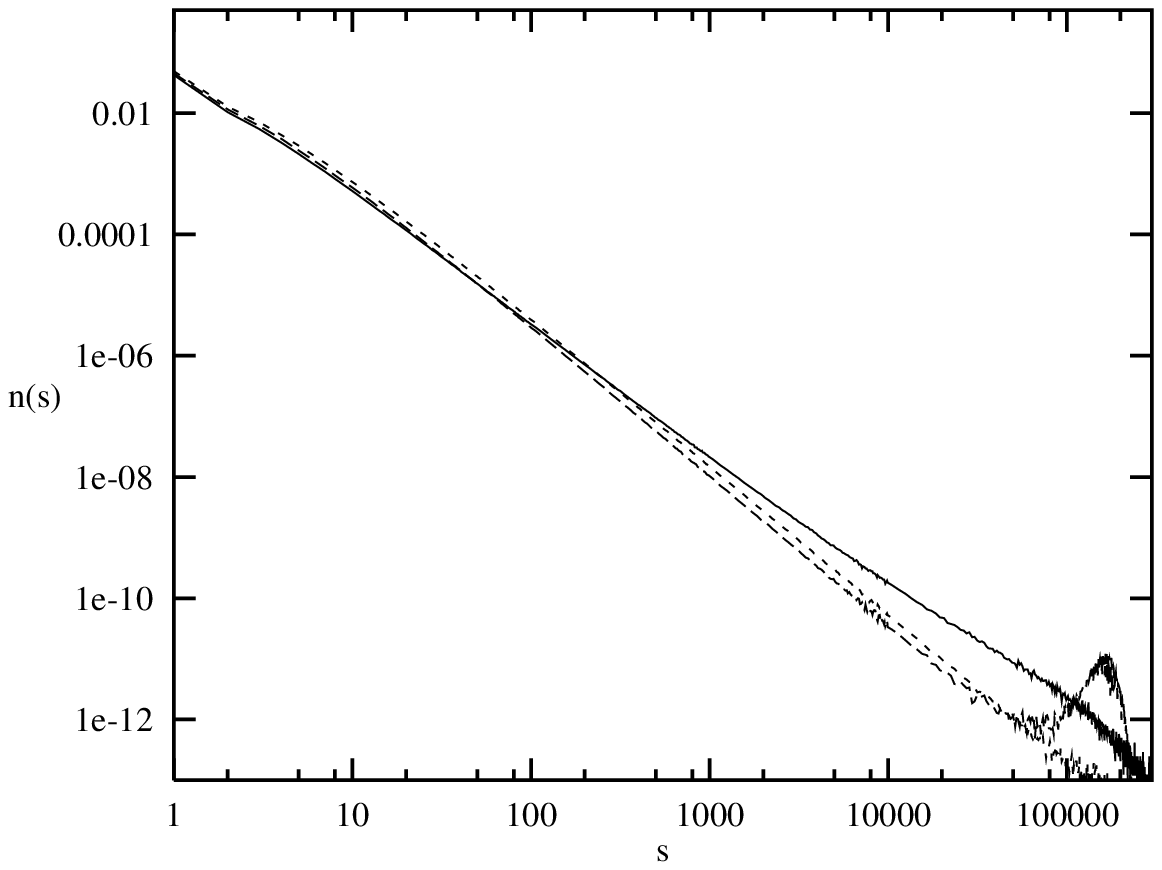}}
\sn
{\par\noindent\figindents
{\bf Fig.\ 6:}
The cluster-size distribution obtained from a simulation of
the full model with $L=16384$ and $f/p = 10^{-4}$ (full line),
the one based on $\pr(\rho)$ (long dashes) and
the result obtained from $\plin(\rho)$ (short dashes).
\par\noindent}
\mn
The determination of the cluster distribution $n(s)$ and $K(y)$
is a complicated combinatorial problem which we solve again
using simulations. One creates configurations
with densities distributed according to the
given $p(\rho)$ and then measures $n(s)$ and $K(y)$.
We have created 100000 configurations on a $512 \times 512$
lattice distributed according to $\pr(\rho)$
and 70000 configurations on a $1024 \times 1024$
lattice for $\plin(\rho)$. Fig.\ 6 shows the cluster-size
distribution $n(s)$ obtained in this manner together with
the one obtained from a simulation of the full model.
For small cluster sizes ($s < 100$), all three distributions
are close to each other. However, at larger $s$ the distributions 
based on a globally given $p(\rho)$ decay faster than the
true $n(s)$. The corresponding exponent is $\tau \approx 2.48$
for the distribution $\pr(\rho)$ and $\tau \approx 2.44$ for
$\plin(\rho)$ -- both much closer to the mean-field
value $\tau = 5/2$ \cite{\CFO} than to the true value \ref{REStau}.
In Fig.\ 6 one also observes a peak in the cluster-size
distribution corresponding to  $\pr(\rho)$ for $1 \cdot 10^5 \le s
\le 2.5 \cdot 10^5 \approx 512^2$, i.e.\ just below the volume
of the system. This is due to the non-vanishing of $\pr(\rho)$
for $\rho > \rho_{{\rm perc}}$ which leads to clusters spanning
a finite (and large) fraction of the system.
\mn
\centerline{\psfig{figure=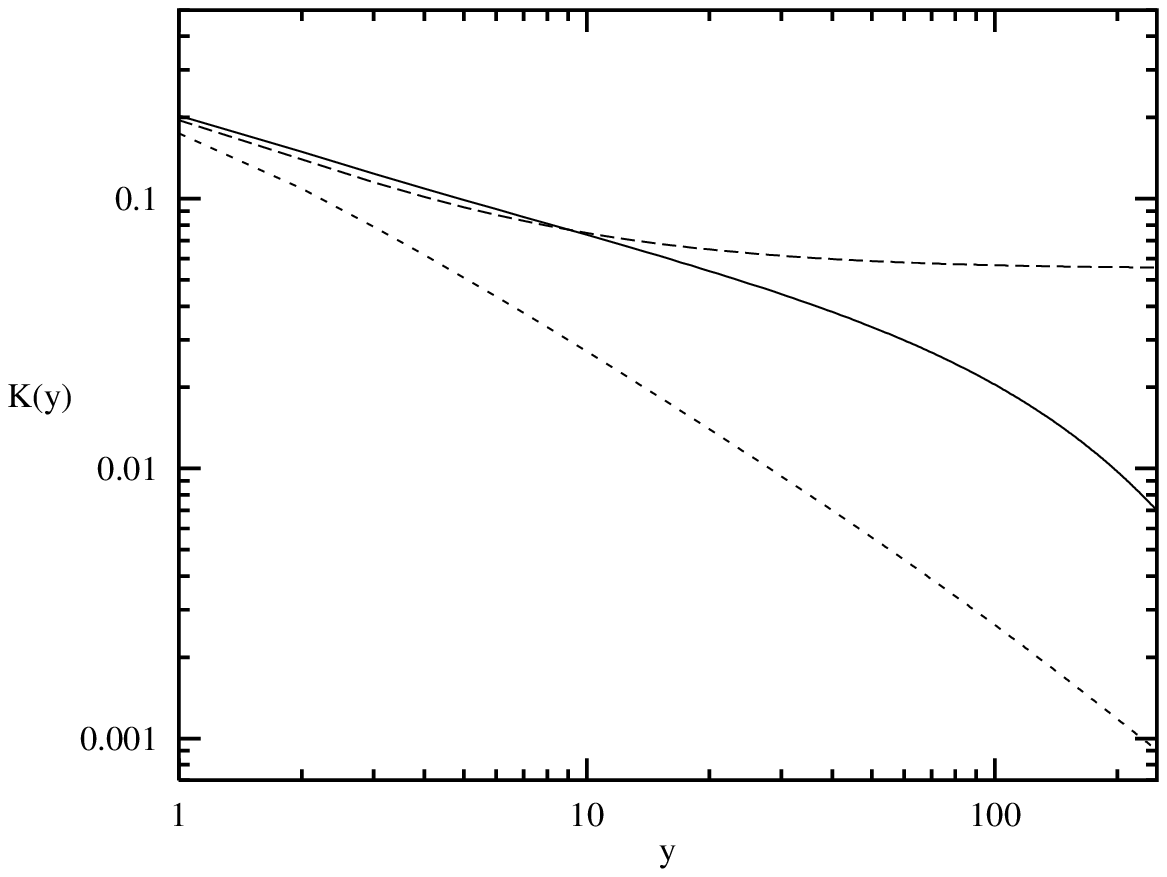}}
\sn
{\par\noindent\figindents
{\bf Fig.\ 7:}
The full line shows the correlation function $K(y)$ obtained
from a simulation of the full model with $L=16384$ and $f/p = 10^{-4}$.
The result for $\pr(\rho)$ is shown by the line with long dashes,
the one obtained from $\plin(\rho)$ is indicated by the shorter dashes.
\par\noindent}
\mn
Fig.\ 7 shows the probability
$K(y)$ to find two trees at distance $y$ inside the same cluster.
Here, the results obtained from the two $p(\rho)$'s deviate more
notably from the result obtained by simulation of the full
model (full line). As noted above, the distribution $\pr(\rho)$
gives rise to percolating clusters
and produces a constant background in $K(y)$ of approximately
$0.0552$. After subtracting this background, one can fit $K(y)$
with \ref{DEFeta} for $\xi_c \approx 180$ and $\eta \approx 0.92$.
This value for $\eta$ is more than twice as large the true one \ref{RESeta}.
For $\plin(\rho)$ one finds good agreement with the form \ref{DEFeta}
for $\xi_c = L/2$ and $\eta \approx 0.95$.
\bn
\subsection{Summary and outlook on global models}
\mn
We have shown that the distributions $\pr$ and $\plin$ lead to
a power law for $n(s)$. One can check that the same is true
for other $p(\rho)$'s that are cut off at $\rho_{{\rm perc}}$
in a way similar to $\plin$. Thus, a power law in $n(s)$ arises
automatically from a description in terms of
global quantities and need not be a signal for criticality
in a conventional sense. However, in all examples for $p(\rho)$
discussed so far, we obtained values for $\tau$ and $\eta$ that
are unsatisfactorily larger than those of the full model.
Nevertheless, a description in terms of $p(\rho)$
can still be forced to work because for two-dimensional critical
percolation one has $\tau \approx 2.055$ and $\eta \approx 0.208$
\cite{\StAh} -- values which are smaller than the ones in Table 2.
So, one can obtain the desired value e.g.\ of $\tau$
by peaking the distribution $p(\rho)$ more prominently just
below $\rho_{{\rm perc}}$. Adjusting just $\tau$ to its correct
value can be expected to also give a reasonably good approximation
for $\eta$. Afterwards, both $\rho_c$ and $a$ may be
tuned to the desired values by adding another peak in $p(\rho)$
for smaller densities (which does not contribute to large clusters
and thus affect the asymptotics of cluster quantities). However,
there does not seem to be a natural way to make these adjustments.
In particular, local densities above $\rho_{{\rm perc}}$ do exist
in the full model which would have to be discarded by hand
in a global model in order to obtain the correct $K(y)$.
\mn
In one dimension the spatial structure becomes irrelevant
at the critical point \cite{\HoPe}. We have seen in this
Section that this can be generalized to the qualitative features
of the critical correlations in two dimensions, but not to the
quantitative details.  In contrast to the one-dimensional case we had
no analytical tools at our disposal and have therefore not been
able to derive the distribution $p(\rho)$ of local densities
explicitly. The introduction
of block-spin variables is reminiscent of real-space renormalization
group ideas and it would be interesting to see if they can be used
to find $p(\rho)$. However, one would have to go beyond the
block-spin renormalization-group study of \cite{\LVZ}. Firstly,
one would have to admit densities different from 0 or 1 for the block-spin
variables, and moreover the dynamics should not be treated
just in mean-field approximation.
\mn
Two-dimensional percolation is believed to be conformally invariant
(see e.g.\ \cite{\LPS}). The globalized models are just ensembles
of percolation problems and should therefore be conformally
invariant as well. It would be interesting to know if also the
stationary state of the full model in two space dimensions
is conformally invariant, even if the standard techniques
of conformal field theory would probably not say much about
quantities like cluster sizes.
\bn
\section{Conclusions}
\mn
In this paper we have again looked at several aspects
of the two-dimensional forest-fire model. Firstly,
we have shown that the two length scales $\xi$ and
$\xi_c$ have different critical exponents. That this
might be possible had been suggested by a study of
the one-dimensional model \cite{\HoPe} which illustrates
that one-dimensional systems can provide useful insights
because of their relative simplicity even if one is
actually interested in higher-dimensional versions.
This result shows that in generic non-equilibrium systems
geometric objects and the usual (occupancy) correlation
functions can behave completely differently.
For an equilibrium system as the Ising model such an observation
was already made some time ago in \cite{\MueK}. In this case,
percolation occurs away from the critical point, i.e.\
the two length scales are so different that they diverge
at different temperatures (see e.g.\ \cite{\CaNa,\StAh}).
\mn
In order to show that $\nu_T \ne \nu$ we had to improve
the error bounds of earlier investigations. As a by-product we
have also improved the accuracy of other critical exponents.
It may be possible that one could still improve the error
bounds by another digit using optimized code on today's most
powerful computers. Historically, Monte-Carlo simulations have
already several times lead to values for the critical exponents that
had to be corrected later on \cite{\DrSchwA,\Henley,\Grassb,\CDrSchE}
and as we have shown here, some of them were still not
treated adequately. Therefore, it may be desirable to perform
yet another independent verification of the results presented
here, but a further increase of accuracy may not be necessary
for this end. 
\mn
The second part of the paper focussed on a globalized model.
We found that one can easily obtain power laws in the cluster-size
distribution by discarding the spatial structure, i.e.\
by making the usual two-point function independent of the spatial
coordinates. This generalizes a result obtained for one dimension
in a previous paper \cite{\HoPe} and is line with the observation
in \cite{\NeSn}, based on a different one-dimensional model,
that one can obtain power laws in clusters or avalanches by
global (`coherent') driving. However, some quantitative
predictions of the globalized model did not work out satisfactorily.
One reason is that the full two-dimensional forest-fire
model has non-trivial two-point functions at least at small distances
which is also reflected by the existence of patches with local densities
above the percolation threshold. In addition, the full model exhibits
many critical exponents that cannot be described by a global model.
\mn
A study of the usual correlation functions would also be desirable
in other models of self-organized criticality where they have not
yet been investigated. This could help to clarify to what extent
the process of self-organization can be regarded as a global
phenomenon. Two-point correlation functions would also be important
quantities to examine in experiments. It would e.g.\ be interesting
to extract the spatial correlation functions of the local heights
and slopes from the experimental data of \cite{\nature}.
\mn
Even for the two-dimensional forest-fire model
there are still many issues we have not looked at, including
e.g.\ finite-size effects. We have only looked at the regime
$f V \gg p$ which is close to the thermodynamic limit. One
could also look at a different limit, namely
$f V \ll p$ where the first-order approximation of \cite{\HoPe}
applies independent of the spatial dimension. In this limit, trees
grow until the lattice is full and after a certain time of rest, lightning
destroys all these trees and the process starts again.
It would be interesting to see how this behaviour crosses over to
the critical behaviour studied here as the volume of the system
is increased, and if finite-size scaling can be observed.
\vskip 0.8 cm
\displayhead{Acknowledgments}
\mn
Useful discussions with S.\ Clar, M.\ Hasenbusch and D.\ Stauffer
are gratefully acknowledged.
The work of A.H.\ has been funded by the Deutsche Forschungsgemeinschaft.
\sectionnumstyle{Alphabetic}
\newsectionnum=0
\vskip 1.6 cm
\appendix{Some exact relaxational modes}
\mn
The time-evolution operator for the forest-fire model in general
dimensions $d$ can be written down along the lines of Section 3 of
\cite{\HoPe}. Then one can obtain {\it exact} excited states using
the same argument as at the end of Section 3 of \cite{\HoPe} (cf.\
in particular eq.\ (3.15) loc.\ cit.).
\mn
The crucial step is to consider a class of configurations that
contain a single cluster of trees and where each empty place is
a neighbour of a tree. The last condition means
that the configuration still consists of a single cluster if a
tree is grown at any of the empty places. Now we consider states
that are built out of such a configuration but have non-zero
momentum $\vec{P} \ne \vec{0}$. Lightning strokes in such
configurations lead to the completely empty system. In states
with $\vec{P} \ne \vec{0}$ the completely empty system is reached
by lightning with coefficients that contain a sum over all
roots of unity such that lightning maps states constructed in
this manner to zero. Thus, the only terms in the image come
from growing a single tree in a configuration. Now, as
in eq.\ (3.15) of \cite{\HoPe} one can write down eigenvalue
equations that are upper triangular in the number of empty
places $N$. The diagonal terms in these equations
are already the eigenvalues
$$\Lambda_N = N p + f/p (V - N) \, .
\label{EigenVal}$$
One must have $N \ge 1$ because the completely full system cannot
be used to build a state with non-zero momentum.
\mn
The number of empty places $N$ is restricted by the above conditions.
For $d=1$ only the two configurations considered in \cite{\HoPe} 
meet the requirement of consisting of a single cluster and
remaining in this class after growing an arbitrary tree. Thus,
$N \le 2$ for $d=1$, and the density $\rho$ of these two
configurations tends to $1$ in the thermodynamic limit.
\mn
For general $d$ note first that the connectedness
of the cluster implies that the trees must form at least
one-dimensional objects. This is most efficiently implemented
by grouping the trees in straight lines. In the hyperplane
perpendicular to these lines, a tree can have $2 d - 2$
empty places as neighbours. Thus, at least one place out of
$2 d - 1$ must be occupied by trees in order to meet the
requirements. This yields $N \ale V ( 1 - 1/(2 d -1))$.
Equivalently, all configurations containing only a single cluster
that retain this property after growing a tree must have
$\rho \ge 1/(2 d -1)$. For $d \le 3$ we can indeed specify configurations
that saturate this lower bound. In $d=1$ this limit is $\rho = 1$
and corresponds to the thermodynamic limit of an almost full system.
In $d=2$ a configuration with $\rho = 1/3$ is given by lines of
trees that are mutually separated by two lines of empty
places. The lower bound $\rho = 1/5$ in $d=3$ is saturated
by arranging the lines in such a way that their mutual position
in the perpendicular plane corresponds to the moves of a pawn
in the game of chess. Of course, the lines have to be connected
by a $d-1$ dimensional object, but this does not change the
value of $\rho$ in the thermodynamic limit.
\vskip 2 cm
\displayhead{References}
\mn
\bibitem{\BTWa} P.\ Bak, C.\ Tang, K.\ Wiesenfeld, {\it Self-Organized
              Criticality: An Explanation of $1/f$ Noise}, Phys.\ Rev.\ Lett.\
              {\bf 59} (1987) 381-384
\bibitem{\BTWb} P.\ Bak, C.\ Tang, K.\ Wiesenfeld, {\it Self-Organized
              Criticality}, Phys Rev.\ {\bf A38} (1988) 364-374
\bibitem{\nature} V.\ Frette, K.\ Christensen, A.\ Malthe-S{\o}renssen,
              J.\ Feder, T.\ J{\o}ssang, P.\ Meakin, {\it Avalanche
              Dynamics in a Pile of Rice}, nature {\bf 379} (1996) 49-52
\bibitem{\BakS} P.\ Bak, K.\ Sneppen, {\it Punctuated Equilibrium and
              Criticality in a Simple Model of Evolution}, Phys.\ Rev.\ Lett.\
              {\bf 71} (1993) 4083-4086
\bibitem{\FBS} P.\ Bak, H.\ Flyvbjerg, K.\ Sneppen, {\it Mean Field Theory for
              a Simple Model of Evolution}, Phys.\ Rev.\ Lett.\ {\bf 71}
              (1993) 4087-4090
\bibitem{\PMB} M.\ Paczuski, S.\ Maslov, P.\ Bak, {\it Avalanche Dynamics in
              Evolution, Growth, and Depinning Models}, Phys.\ Rev.\ {\bf
              E53} (1996) 414-443
\bibitem{\HenOld} C.L.\ Henley, {\it Self-Organized Percolation: A Simpler
              Model}, Bull.\ Am.\ Phys.\ Soc.\ {\bf 34} (1989) 838
\bibitem{\DrSchwA} B.\ Drossel, F.\ Schwabl, {\it Self-Organized Critical
              Forest-Fire Model}, Phys.\ Rev.\ Lett.\ {\bf 69} (1992)
              1629-1632
\bibitem{\BCT} P.\ Bak, K.\ Chen, C.\ Tang, {\it A Forest-Fire Model and Some
              Thoughts on Turbulence}, Phys.\ Lett.\ {\bf A147} (1990)
              297-300
\bibitem{\GrKa} P.\ Grassberger, H.\ Kantz, {\it On a Forest Fire Model with
              Supposed Self-Organized Criticality}, J.\ Stat.\ Phys.\ {\bf
              63} (1991) 685-700
\bibitem{\DMS} W.K.\ Mo{\ss}ner, B.\ Drossel, F.\ Schwabl, {\it Computer
              Simulations of the Forest-Fire Model}, Physica {\bf A190}
              (1992) 205-217
\bibitem{\Henley} C.L.\ Henley, {\it Statics of a ``Self-Organized''
              Percolation Model}, Phys.\ Rev.\ Lett.\ {\bf 71} (1993) 2741-2744
\bibitem{\Grassb} P.\ Grassberger, {\it On a Self-Organized Critical
              Forest-Fire Model},
              J.\ Phys.\ A: Math.\ Gen.\ {\bf 26} (1993) 2081-2089
\bibitem{\CDrSchE} S.\ Clar, B.\ Drossel, F.\ Schwabl, {\it Scaling Laws and
              Simulation Results for the Self-Organized Critical Forest-Fire
              Model}, Phys.\ Rev.\ {\bf E50} (1994) 1009-1018
\bibitem{\HoPe} A.\ Honecker, I.\ Peschel,
              {\it Critical Properties of the One-Dimensional
              Forest-Fire Model}, Physica {\bf A229} (1996) 478-500
\bibitem{\NeSn} M.E.J.\ Newman, K.\ Sneppen, {\it Avalanches, Scaling and
              Coherent Noise}, preprint cond-mat/9606066, CTC96TR237
\bibitem{\BJW} J.\ de Boer, A.D.\ Jackson, T.\ Wettig, {\it Criticality in
              Simple Models of Evolution}, Phys.\ Rev.\ {\bf E51} (1995)
              1059-1074
\bibitem{\LVZ} V.\ Loreto, A.\ Vespignani, S.\ Zapperi, {\it Renormalization
              Scheme for Forest-Fire Models}, J.\ Phys.\ A: Math.\ Gen.\ {\bf
              29} (1996) 2981-3004
\bibitem{\WWW} \href{http://www.physik.fu-berlin.de/~ag-peschel/software/forest2d.html}
              {http://www.physik.fu-berlin.de/\~{}ag-peschel/software/forest2d.html}
\bibitem{\ClarP} S.\ Clar, private communication
\bibitem{\CFO} K.\ Christensen, H.\ Flyvbjerg, Z.\ Olami, {\it Self-Organized
              Critical Forest-Fire Model: Mean-Field Theory and Simulation
              Results in 1 to 6 Dimensions}, Phys.\ Rev.\ Lett.\ {\bf 71}
              (1993) 2737-2740
\bibitem{\CDrSchw} B.\ Drossel, S.\ Clar, F.\ Schwabl, {\it Exact Results for
              the One-Dimensional Self-Organized Critical Forest-Fire Model},
              Phys.\ Rev.\ Lett.\ {\bf 71} (1993) 3739-3742
\bibitem{\StAh} D.\ Stauffer, A.\ Aharony, {\it Perkolationstheorie,
              Eine Einf\"uhrung}, VCH Weinheim (1995) (German translation of
              {\it Introduction to Percolation Theory}, Taylor \& Francis
              London (1992))
\bibitem{\DrSchF} B.\ Drossel, F.\ Schwabl, {\it Self Organization in a
              Forest-Fire Model},  Fractals {\bf 1} (1993) 1022-1029
\bibitem{\Drossel} B.\ Drossel, {\it Strukturbildung in offenen Systemen},
              Ph.D.\ thesis, M\"unchen (1994)
\bibitem{\LPS} R.\ Langlands, P.\ Pouliot, Y.\ Saint-Aubin, {\it Conformal
              Invariance in Two-Dimensional Percolation}, Bull.\ Am.\ Math.\
              Soc.\ {\bf 30} (1994) 1-61
\bibitem{\MueK} H.\ M\"uller-Krumbhaar, {\it The Droplet Model in Three
              Dimensions: Monte Carlo Calculation Results}, Phys.\ Lett.\
              {\bf A48} (1974) 459-460
\bibitem{\CaNa} J.L.\ Cambier, M.\ Nauenberg, {\it Distribution of Fractal
              Clusters and Scaling in the Ising Model}, Phys.\ Rev.\ {\bf
              B34} (1986) 8071-8079
\vfill
\end

%% file: ffm_macros.tex


\catcode`\@=11


\message{Loading a modification of the jyTeX macros...}

\message{modifications to plain.tex,}


\def\newcount{\alloc@0\count\countdef\insc@unt}
\def\newdimen{\alloc@1\dimen\dimendef\insc@unt}
\def\newskip{\alloc@2\skip\skipdef\insc@unt}
\def\newmuskip{\alloc@3\muskip\muskipdef\@cclvi}
\def\newtoks{\alloc@5\toks\toksdef\@cclvi}
\def\newhelp#1#2{\newtoks#1\global#1\expandafter{\csname#2\endcsname}}
\def\newread{\alloc@6\read\chardef\sixt@@n}
\def\newwrite{\alloc@7\write\chardef\sixt@@n}
\def\newfam{\alloc@8\fam\chardef\sixt@@n}
\def\newinsert#1{\global\advance\insc@unt by\m@ne
     \ch@ck0\insc@unt\count
     \ch@ck1\insc@unt\dimen
     \ch@ck2\insc@unt\skip
     \ch@ck4\insc@unt\box
     \allocationnumber=\insc@unt
     \global\chardef#1=\allocationnumber
     \wlog{\string#1=\string\insert\the\allocationnumber}}
\def\newif#1{\count@\escapechar \escapechar\m@ne
     \expandafter\expandafter\expandafter
          \xdef\@if#1{true}{\let\noexpand#1=\noexpand\iftrue}%
     \expandafter\expandafter\expandafter
          \xdef\@if#1{false}{\let\noexpand#1=\noexpand\iffalse}%
     \global\@if#1{false}\escapechar=\count@}


\newlinechar=`\^^J
\overfullrule=0pt

\message{hacks,}


\toksdef\toks@i=1
\toksdef\toks@ii=2


\def\TeX{T\kern-.1667em \lower.5ex \hbox{E}\kern-.125em X\null}
\def\jyTeX{{\leavevmode
     \raise.587ex \hbox{\it\j}\kern-.1em \lower.048ex \hbox{\it y}\kern-.12em
     \TeX}}

\let\then=\iftrue
\def\ifnoarg#1\then{\def\hack@{#1}\ifx\hack@\empty}
\def\ifundefined#1\then{%
     \expandafter\ifx\csname\expandafter\blank\string#1\endcsname\relax}
\def\useif#1\then{\csname#1\endcsname}
\def\usename#1{\csname#1\endcsname}
\def\useafter#1#2{\expandafter#1\csname#2\endcsname}

\long\def\loop#1\repeat{\def\@iterate{#1\expandafter\@iterate\fi}\@iterate
     \let\@iterate=\relax}

\let\TeXend=\end
\def\begin#1{\begingroup\def\@@blockname{#1}\usename{begin#1}}
\def\End#1{\usename{end#1}\def\hack@{#1}%
     \ifx\@@blockname\hack@
          \endgroup
     \else\err@badgroup\hack@\@@blockname
     \fi}
\def\@@blockname{}

\def\defaultoption[#1]#2{%
     \def\hack@{\ifx\hack@ii[\toks@={#2}\else\toks@={#2[#1]}\fi\the\toks@}%
     \futurelet\hack@ii\hack@}

\def\markup#1{\let\@@marksf=\empty
     \ifhmode\edef\@@marksf{\spacefactor=\the\spacefactor\relax}\/\fi
     ${}^{\hbox{\subscriptfonts#1}}$\@@marksf}


\newtoks\shortyear
\newtoks\militaryhour
\newtoks\standardhour
\newtoks\minute
\newtoks\amorpm

\def\settime{\count@=\time\divide\count@ by60
     \militaryhour=\expandafter{\number\count@}%
     {\multiply\count@ by-60 \advance\count@ by\time
          \xdef\hack@{\ifnum\count@<10 0\fi\number\count@}}%
     \minute=\expandafter{\hack@}%
     \ifnum\count@<12
          \amorpm={am}
     \else\amorpm={pm}
          \ifnum\count@>12 \advance\count@ by-12 \fi
     \fi
     \standardhour=\expandafter{\number\count@}%
     \def\hack@19##1##2{\shortyear={##1##2}}%
          \expandafter\hack@\the\year}

\def\monthword#1{%
     \ifcase#1
          $\bullet$\err@badcountervalue{monthword}%
          \or January\or February\or March\or April\or May\or June%
          \or July\or August\or September\or October\or November\or December%
     \else$\bullet$\err@badcountervalue{monthword}%
     \fi}

\def\monthabbr#1{%
     \ifcase#1
          $\bullet$\err@badcountervalue{monthabbr}%
          \or Jan\or Feb\or Mar\or Apr\or May\or Jun%
          \or Jul\or Aug\or Sep\or Oct\or Nov\or Dec%
     \else$\bullet$\err@badcountervalue{monthabbr}%
     \fi}

\def\militarytime{\the\militaryhour:\the\minute}
\def\standardtime{\the\standardhour:\the\minute}


\def\@setnumstyle#1#2{\expandafter\global\expandafter\expandafter
     \expandafter\let\expandafter\expandafter
     \csname @\expandafter\blank\string#1style\endcsname
     \csname#2\endcsname}
\def\numstyle#1{\usename{@\expandafter\blank\string#1style}#1}
\def\ifblank#1\then{\useafter\ifx{@\expandafter\blank\string#1}\blank}

\def\blank#1{}

\def\Roman#1{\expandafter\uppercase\expandafter{\romannumeral#1}}
\def\alphabetic#1{%
     \ifcase#1
          $\bullet$\err@badcountervalue{alphabetic}%
          \or a\or b\or c\or d\or e\or f\or g\or h\or i\or j\or k\or l\or m%
          \or n\or o\or p\or q\or r\or s\or t\or u\or v\or w\or x\or y\or z%
     \else$\bullet$\err@badcountervalue{alphabetic}%
     \fi}
\def\Alphabetic#1{\expandafter\uppercase\expandafter{\alphabetic{#1}}}
\def\symbols#1{%
     \ifcase#1
          $\bullet$\err@badcountervalue{symbols}%
          \or*\or\dag\or\ddag\or\S\or$\|$%
          \or**\or\dag\dag\or\ddag\ddag\or\S\S\or$\|\|$%
     \else$\bullet$\err@badcountervalue{symbols}%
     \fi}


\catcode`\^^?=13 \def^^?{\relax}

\def\trimleading#1\to#2{\edef#2{#1}%
     \expandafter\@trimleading\expandafter#2#2^^?^^?}
\def\@trimleading#1#2#3^^?{\ifx#2^^?\def#1{}\else\def#1{#2#3}\fi}

\def\trimtrailing#1\to#2{\edef#2{#1}%
     \expandafter\@trimtrailing\expandafter#2#2^^? ^^?\relax}
\def\@trimtrailing#1#2 ^^?#3{\ifx#3\relax\toks@={}%
     \else\def#1{#2}\toks@={\trimtrailing#1\to#1}\fi
     \the\toks@}

\def\trim#1\to#2{\trimleading#1\to#2\trimtrailing#2\to#2}

\catcode`\^^?=15


\long\def\additemL#1\to#2{\toks@={\^^\{#1}}\toks@ii=\expandafter{#2}%
     \xdef#2{\the\toks@\the\toks@ii}}

\long\def\additemR#1\to#2{\toks@={\^^\{#1}}\toks@ii=\expandafter{#2}%
     \xdef#2{\the\toks@ii\the\toks@}}

\def\getitemL#1\to#2{\expandafter\@getitemL#1\hack@#1#2}
\def\@getitemL\^^\#1#2\hack@#3#4{\def#4{#1}\def#3{#2}}


\newskip\headskip
\newskip\footskip

\message{document layout,}

\newif\ifdraft
\def\draft{\drafttrue\leftmargin=.5in \overfullrule=5pt }


\newskip\abovechapterskip
\newskip\belowchapterskip
\newskip\abovesectionskip
\newskip\belowsectionskip
\newskip\abovesubsectionskip
\newskip\belowsubsectionskip

\def\chapterstyle#1{\global\expandafter\let\expandafter\@chapterstyle
     \csname#1text\endcsname}
\def\sectionstyle#1{\global\expandafter\let\expandafter\@sectionstyle
     \csname#1text\endcsname}
\def\subsectionstyle#1{\global\expandafter\let\expandafter\@subsectionstyle
     \csname#1text\endcsname}

\def\CHapter#1{%
     \ifdim\lastskip=17sp \else\chapterbreak\vskip\abovechapterskip\fi
     \@chapterstyle{\ifblank\chapternumstyle\then
          \else\newchapternum=\next\chapternumformat\ \fi#1}%
     \nobreak\vskip\belowchapterskip\vskip17sp }

\def\Section#1{%
     \ifdim\lastskip=17sp \else\sectionbreak\vskip\abovesectionskip\fi
     \@sectionstyle{\ifblank\sectionnumstyle\then
          \else\newsectionnum=\next\sectionnumformat\ \fi#1}%
     \nobreak\vskip\belowsectionskip\vskip17sp }

\def\Subsection#1{%
     \ifdim\lastskip=17sp \else\subsectionbreak\vskip\abovesubsectionskip\fi
     \@subsectionstyle{\ifblank\subsectionnumstyle\then
          \else\newsubsectionnum=\next\subsectionnumformat\ \fi#1}%
     \nobreak\vskip\belowsubsectionskip\vskip17sp }


\newtoks\everybye \everybye={\par\vfil}
\outer\def\bye{\the\everybye
     \footnotecheck
     \prelabelcheck
     \streamcheck
     \supereject
     \TeXend}

\message{labels,}

\let\@@labeldef=\xdef
\newif\if@labelfile
\newwrite\@labelfile
\let\@prelabellist=\empty

\def\Label#1#2{\trim#1\to\@@labarg\edef\@@labtext{#2}%
     \edef\@@labname{lab@\@@labarg}%
     \useafter\ifundefined\@@labname\then\else\@yeslab\fi
     \useafter\@@labeldef\@@labname{#2}%
     \ifstreaming
          \expandafter\toks@\expandafter\expandafter\expandafter
               {\csname\@@labname\endcsname}%
          \immediate\write\streamout{\noexpand\Label{\@@labarg}{\the\toks@}}%
     \fi}
\def\@yeslab{%
     \useafter\ifundefined{if\@@labname}\then
          \err@labelredef\@@labarg
     \else\useif{if\@@labname}\then
               \err@labelredef\@@labarg
          \else\global\usename{\@@labname true}%
               \useafter\ifundefined{pre\@@labname}\then
               \else\useafter\ifx{pre\@@labname}\@@labtext
                    \else\err@badlabelmatch\@@labarg
                    \fi
               \fi
               \if@labelfile
               \else\global\@labelfiletrue
                    \immediate\write\sixt@@n{--> Creating file \jobname.lab}%
                    \immediate\openout\@labelfile=\jobname.lab
               \fi
               \immediate\write\@labelfile
                    {\noexpand\prelabel{\@@labarg}{\@@labtext}}%
          \fi
     \fi}

\def\putlab#1{\trim#1\to\@@labarg\edef\@@labname{lab@\@@labarg}%
     \useafter\ifundefined\@@labname\then\@nolab\else\usename\@@labname\fi}
\def\@nolab{%
     \useafter\ifundefined{pre\@@labname}\then
          \undefinedlabelformat
          \err@needlabel\@@labarg
          \useafter\xdef\@@labname{\undefinedlabelformat}%
     \else\usename{pre\@@labname}%
          \useafter\xdef\@@labname{\usename{pre\@@labname}}%
     \fi
     \useafter\newif{if\@@labname}%
     \expandafter\additemR\@@labarg\to\@prelabellist}

\def\prelabel#1{\useafter\gdef{prelab@#1}}

\def\ifundefinedlabel#1\then{%
     \expandafter\ifx\csname lab@#1\endcsname\relax}
\def\useiflab#1\then{\csname iflab@#1\endcsname}

\def\prelabelcheck{{%
     \def\^^\##1{\useiflab{##1}\then\else\err@undefinedlabel{##1}\fi}%
     \@prelabellist}}

\message{equation numbering,}

\newcount\chapternum
\newcount\sectionnum
\newcount\subsectionnum
\newcount\equationnum
\newcount\subequationnum
\newcount\figurenum
\newcount\subfigurenum
\newcount\tablenum
\newcount\subtablenum
\newcount\defnum
\newcount\subdefnum
\newcount\thmnum
\newcount\subthmnum
\newcount\lemnum
\newcount\sublemnum

\newif\if@subeqncount
\newif\if@subfigcount
\newif\if@subtblcount
\newif\if@subdefcount
\newif\if@subthmcount
\newif\if@sublemcount

\def\newchapternum{\newsectionnum=\z@\@resetnum\chapternum}
\def\newsectionnum{\newsubsectionnum=\z@\@resetnum\sectionnum}
\def\newsubsectionnum{\newequationnum=\z@\newfigurenum=\z@\newtablenum=\z@
     \newdefnum=\z@\newthmnum=\z@\newlemnum=\z@
     \@resetnum\subsectionnum}
\def\newequationnum{\newsubequationnum=\z@\@resetnum\equationnum}
\def\newsubequationnum{\@resetnum\subequationnum}
\def\newfigurenum{\newsubfigurenum=\z@\@resetnum\figurenum}
\def\newsubfigurenum{\@resetnum\subfigurenum}
\def\newtablenum{\newsubtablenum=\z@\@resetnum\tablenum}
\def\newsubtablenum{\@resetnum\subtablenum}
\def\newdefnum{\newsubdefnum=\z@\@resetnum\defnum}
\def\newsubdefnum{\@resetnum\subdefnum}
\def\newthmnum{\newsubthmnum=\z@\@resetnum\thmnum}
\def\newsubthmnum{\@resetnum\subthmnum}
\def\newlemnum{\newsublemnum=\z@\@resetnum\lemnum}
\def\newsublemnum{\@resetnum\sublemnum}

\def\@resetnum#1{\global\advance#1by1 \edef\next{\the#1\relax}\global#1}

\newchapternum=0

\def\chapternumstyle#1{\@setnumstyle\chapternum{#1}}
\def\sectionnumstyle#1{\@setnumstyle\sectionnum{#1}}
\def\subsectionnumstyle#1{\@setnumstyle\subsectionnum{#1}}
\def\equationnumstyle#1{\@setnumstyle\equationnum{#1}}
\def\subequationnumstyle#1{\@setnumstyle\subequationnum{#1}%
     \ifblank\subequationnumstyle\then\global\@subeqncountfalse\fi
     \ignorespaces}
\def\figurenumstyle#1{\@setnumstyle\figurenum{#1}}
\def\subfigurenumstyle#1{\@setnumstyle\subfigurenum{#1}%
     \ifblank\subfigurenumstyle\then\global\@subfigcountfalse\fi
     \ignorespaces}
\def\tablenumstyle#1{\@setnumstyle\tablenum{#1}}
\def\subtablenumstyle#1{\@setnumstyle\subtablenum{#1}%
     \ifblank\subtablenumstyle\then\global\@subtblcountfalse\fi
     \ignorespaces}
\def\defnumstyle#1{\@setnumstyle\defnum{#1}}
\def\subdefnumstyle#1{\@setnumstyle\subdefnum{#1}%
     \ifblank\subdefnumstyle\then\global\@subdefcountfalse\fi
     \ignorespaces}
\def\thmnumstyle#1{\@setnumstyle\thmnum{#1}}
\def\subthmnumstyle#1{\@setnumstyle\subthmnum{#1}%
     \ifblank\subthmnumstyle\then\global\@subthmcountfalse\fi
     \ignorespaces}
\def\lemnumstyle#1{\@setnumstyle\lemnum{#1}}
\def\sublemnumstyle#1{\@setnumstyle\sublemnum{#1}%
     \ifblank\sublemnumstyle\then\global\@sublemcountfalse\fi
     \ignorespaces}

\def\heqnlabel{\newequationnum=\next
          \ifblank\subequationnumstyle\then
          \else\global\@subeqncounttrue
               \newsubequationnum=\@ne
          \fi}

\def\eqnlabel#1{%
     \if@subeqncount
          \newsubequationnum=\next
     \else\heqnlabel
     \fi
     \Label{#1}{\puteqnformat}(\puteqn{#1})%
     \ifdraft\rlap{\hskip.1in{\tt#1}}\fi}

\let\puteqn=\putlab

\def\putequation#1{\useafter\ifundefined{eqn@#1}\then
     \err@undefinedeqn{#1}\else\usename{eqn@#1}\fi}

\def\eqnseriesstyle#1{\gdef\@eqnseriesstyle{#1}}
\def\begineqnseries{\subequationnumstyle{\@eqnseriesstyle}%
     \defaultoption[]\@begineqnseries}
\def\@begineqnseries[#1]{\edef\@@eqnname{#1}}
\def\endeqnseries{\subequationnumstyle{blank}%
     \expandafter\ifnoarg\@@eqnname\then
     \else\Label\@@eqnname{\puteqnformat}%
     \fi
     \aftergroup\ignorespaces}

\def\figlabel#1{%
     \if@subfigcount
          \newsubfigurenum=\next
     \else\newfigurenum=\next
          \ifblank\subfigurenumstyle\then
          \else\global\@subfigcounttrue
               \newsubfigurenum=\@ne
          \fi
     \fi
     \Label{#1}{\putfigformat}\putfig{#1}%
   }

\let\putfig=\putlab

\def\figseriesstyle#1{\gdef\@figseriesstyle{#1}}
\def\beginfigseries{\subfigurenumstyle{\@figseriesstyle}%
     \defaultoption[]\@beginfigseries}
\def\@beginfigseries[#1]{\edef\@@figname{#1}}
\def\endfigseries{\subfigurenumstyle{blank}%
     \expandafter\ifnoarg\@@figname\then
     \else\Label\@@figname{\putfigformat}%
     \fi
     \aftergroup\ignorespaces}

\def\tbllabel#1{%
     \if@subtblcount
          \newsubtablenum=\next
     \else\newtablenum=\next
          \ifblank\subtablenumstyle\then
          \else\global\@subtblcounttrue
               \newsubtablenum=\@ne
          \fi
     \fi
     \Label{#1}{\puttblformat}\puttbl{#1}%
}

\let\puttbl=\putlab

\def\tblseriesstyle#1{\gdef\@tblseriesstyle{#1}}
\def\begintblseries{\subtablenumstyle{\@tblseriesstyle}%
     \defaultoption[]\@begintblseries}
\def\@begintblseries[#1]{\edef\@@tblname{#1}}
\def\endtblseries{\subtablenumstyle{blank}%
     \expandafter\ifnoarg\@@tblname\then
     \else\Label\@@tblname{\puttblformat}%
     \fi
     \aftergroup\ignorespaces}


\def\deflab#1{%
     \if@subdefcount
          \newsubdefnum=\next
     \else\newdefnum=\next
          \ifblank\subdefnumstyle\then
          \else\global\@subdefcounttrue
               \newsubdefnum=\@ne
          \fi
     \fi
     \Label{#1}{\putdefformat}\refdef{#1}%
}

\let\refdef=\putlab

\def\defseriesstyle#1{\gdef\@defseriesstyle{#1}}
\def\begindefseries{\subtablenumstyle{\@defseriesstyle}%
     \defaultoption[]\@begindefseries}
\def\@begindefseries[#1]{\edef\@@defname{#1}}
\def\enddefseries{\subdefnumstyle{blank}%
     \expandafter\ifnoarg\@@defname\then
     \else\Label\@@defname{\putdefformat}%
     \fi
     \aftergroup\ignorespaces}

\def\thmlab#1{%
     \if@subthmcount
          \newsubthmnum=\next
     \else\newthmnum=\next
          \ifblank\subthmnumstyle\then
          \else\global\@subthmcounttrue
               \newsubthmnum=\@ne
          \fi
     \fi
     \Label{#1}{\putthmformat}\refthm{#1}%
}

\let\refthm=\putlab

\def\thmseriesstyle#1{\gdef\@thmseriesstyle{#1}}
\def\beginthmseries{\subthmnumstyle{\@thmseriesstyle}%
     \defaultoption[]\@beginthmseries}
\def\@beginthmseries[#1]{\edef\@@thmname{#1}}
\def\endthmseries{\subthmstyle{blank}%
     \expandafter\ifnoarg\@@thmname\then
     \else\Label\@@thmname{\putthmformat}%
     \fi
     \aftergroup\ignorespaces}

\def\lemlab#1{%
     \if@sublemcount
          \newsublemnum=\next
     \else\newlemnum=\next
          \ifblank\sublemnumstyle\then
          \else\global\@sublemcounttrue
               \newsublemnum=\@ne
          \fi
     \fi
     \Label{#1}{\putlemformat}\reflem{#1}%
}

\let\reflem=\putlab

\def\lemseriesstyle#1{\gdef\@lemseriesstyle{#1}}
\def\beginlemseries{\sublemnumstyle{\@lemseriesstyle}%
     \defaultoption[]\@beginlemseries}
\def\@beginlemseries[#1]{\edef\@@lemname{#1}}
\def\endlemseries{\sublemnumstyle{blank}%
     \expandafter\ifnoarg\@@lemname\then
     \else\Label\@@lemname{\putlemformat}%
     \fi
     \aftergroup\ignorespaces}

\message{reference numbering,}

\newcount\referencenum \referencenum=0
\newcount\@@prerefcount \@@prerefcount=0
\newcount\@@thisref
\newcount\@@lastref
\newcount\@@loopref
\newcount\@@refseq
\newdimen\refnumindent
\let\@undefreflist=\empty

\def\referencenumstyle#1{\@setnumstyle\referencenum{#1}}

\def\referencestyle#1{\usename{@ref#1}}

\def\@refsequential{%
     \gdef\@refpredef##1{\global\advance\referencenum by\@ne
          \let\^^\=0\Label{##1}{\^^\{\the\referencenum}}%
          \useafter\gdef{ref@\the\referencenum}{{##1}{\undefinedlabelformat}}}%
     \gdef\@reference##1##2{%
          \ifundefinedlabel##1\then
          \else\def\^^\####1{\global\@@thisref=####1\relax}\putlab{##1}%
               \useafter\gdef{ref@\the\@@thisref}{{##1}{##2}}%
          \fi}%
     \gdef\endputreferences{%
          \loop\ifnum\@@loopref<\referencenum
                    \advance\@@loopref by\@ne
                    \expandafter\expandafter\expandafter\@printreference
                         \csname ref@\the\@@loopref\endcsname
          \repeat
          \par}}

\def\@refpreordered{%
     \gdef\@refpredef##1{\global\advance\referencenum by\@ne
          \additemR##1\to\@undefreflist}%
     \gdef\@reference##1##2{%
          \ifundefinedlabel##1\then
          \else\global\advance\@@loopref by\@ne
               {\let\^^\=0\Label{##1}{\^^\{\the\@@loopref}}}%
               \@printreference{##1}{##2}%
          \fi}
     \gdef\endputreferences{%
          \def\^^\####1{\useiflab{####1}\then
               \else\reference{####1}{\undefinedlabelformat}\fi}%
          \@undefreflist
          \par}}

\def\beginprereferences{\par
     \def\reference##1##2{\global\advance\referencenum by1\@ne
          \let\^^\=0\Label{##1}{\^^\{\the\referencenum}}%
          \useafter\gdef{ref@\the\referencenum}{{##1}{##2}}}}
\def\endprereferences{\global\@@prerefcount=\the\referencenum\par}

\def\beginputreferences{\par
     \refnumindent=\z@\@@loopref=\z@
     \loop\ifnum\@@loopref<\referencenum
               \advance\@@loopref by\@ne
               \setbox\z@=\hbox{\referencenum=\@@loopref
                    \referencenumformat\enskip}%
               \ifdim\wd\z@>\refnumindent\refnumindent=\wd\z@\fi
     \repeat
     \putreferenceformat
     \@@loopref=\z@
     \loop\ifnum\@@loopref<\@@prerefcount
               \advance\@@loopref by\@ne
               \expandafter\expandafter\expandafter\@printreference
                    \csname ref@\the\@@loopref\endcsname
     \repeat
     \let\reference=\@reference}

\def\@printreference#1#2{\ifx#2\undefinedlabelformat\err@undefinedref{#1}\fi
     \noindent\ifdraft\rlap{\hskip\hsize\hskip.1in \tt#1}\fi
     \llap{\referencenum=\@@loopref\referencenumformat\enskip}#2\par}

\def\reference#1#2{{\par\refnumindent=\z@\putreferenceformat\noindent#2\par}}

\def\putref#1{\trim#1\to\@@refarg
     \expandafter\ifnoarg\@@refarg\then
          \toks@={\relax}%
     \else\@@lastref=-\@m\def\@@refsep{}\def\@more{\@nextref}%
          \toks@={\@nextref#1,,}%
     \fi\the\toks@}
\def\@nextref#1,{\trim#1\to\@@refarg
     \expandafter\ifnoarg\@@refarg\then
          \let\@more=\relax
     \else\ifundefinedlabel\@@refarg\then
               \expandafter\@refpredef\expandafter{\@@refarg}%
          \fi
          \def\^^\##1{\global\@@thisref=##1\relax}%
          \global\@@thisref=\m@ne
          \setbox\z@=\hbox{\putlab\@@refarg}%
     \fi
     \advance\@@lastref by\@ne
     \ifnum\@@lastref=\@@thisref\advance\@@refseq by\@ne\else\@@refseq=\@ne\fi
     \ifnum\@@lastref<\z@
     \else\ifnum\@@refseq<\thr@@
               \@@refsep\def\@@refsep{,}%
               \ifnum\@@lastref>\z@
                    \advance\@@lastref by\m@ne
                    {\referencenum=\@@lastref\putrefformat}%
               \else\undefinedlabelformat
               \fi
          \else\def\@@refsep{--}%
          \fi
     \fi
     \@@lastref=\@@thisref
     \@more}

\message{streaming,}

\newif\ifstreaming

\def\streamto{\defaultoption[\jobname]\@streamto}
\def\@streamto[#1]{\global\streamingtrue
     \immediate\write\sixt@@n{--> Streaming to #1.str}%
     \newwrite\streamout\immediate\openout\streamout=#1.str }

\def\streamfrom{\defaultoption[\jobname]\@streamfrom}
\def\@streamfrom[#1]{\newread\streamin\openin\streamin=#1.str
     \ifeof\streamin
          \expandafter\err@nostream\expandafter{#1.str}%
     \else\immediate\write\sixt@@n{--> Streaming from #1.str}%
          \let\@@labeldef=\gdef
          \ifstreaming
               \edef\@elc{\endlinechar=\the\endlinechar}%
               \endlinechar=\m@ne
               \loop\read\streamin to\@@scratcha
                    \ifeof\streamin
                         \streamingfalse
                    \else\toks@=\expandafter{\@@scratcha}%
                         \immediate\write\streamout{\the\toks@}%
                    \fi
                    \ifstreaming
               \repeat
               \@elc
               \input #1.str
               \streamingtrue
          \else\input #1.str
          \fi
          \let\@@labeldef=\xdef
     \fi}

\def\streamcheck{\ifstreaming
     \immediate\write\streamout{\pagenum=\the\pagenum}%
     \immediate\write\streamout{\footnotenum=\the\footnotenum}%
     \immediate\write\streamout{\referencenum=\the\referencenum}%
     \immediate\write\streamout{\chapternum=\the\chapternum}%
     \immediate\write\streamout{\sectionnum=\the\sectionnum}%
     \immediate\write\streamout{\subsectionnum=\the\subsectionnum}%
     \immediate\write\streamout{\equationnum=\the\equationnum}%
     \immediate\write\streamout{\subequationnum=\the\subequationnum}%
     \immediate\write\streamout{\figurenum=\the\figurenum}%
     \immediate\write\streamout{\subfigurenum=\the\subfigurenum}%
     \immediate\write\streamout{\tablenum=\the\tablenum}%
     \immediate\write\streamout{\subtablenum=\the\subtablenum}%
     \immediate\closeout\streamout
     \fi}


\def\err@badtypesize{%
     \errhelp={The limited availability of certain fonts requires^^J%
          that the base type size be 10pt, 12pt, or 14pt.^^J}%
     \errmessage{--> Illegal base type size}}

\def\err@badsizechange{\immediate\write\sixt@@n
     {--> Size change not allowed in math mode, ignored}}

\def\err@sizetoolarge#1{\immediate\write\sixt@@n
     {--> \noexpand#1 too big, substituting HUGE}}

\def\err@sizenotavailable#1{\immediate\write\sixt@@n
     {--> Size not available, \noexpand#1 ignored}}

\def\err@fontnotavailable#1{\immediate\write\sixt@@n
     {--> Font not available, \noexpand#1 ignored}}

\def\err@sltoit{\immediate\write\sixt@@n
     {--> Style \noexpand\sl not available, substituting \noexpand\it}%
     \it}

\def\err@bfstobf{\immediate\write\sixt@@n
     {--> Style \noexpand\bfs not available, substituting \noexpand\bf}%
     \bf}

\def\err@badgroup#1#2{%
     \errhelp={The block you have just tried to close was not the one^^J%
          most recently opened.^^J}%
     \errmessage{--> \noexpand\End{#1} doesn't match \noexpand\begin{#2}}}

\def\err@badcountervalue#1{\immediate\write\sixt@@n
     {--> Counter (#1) out of bounds}}

\def\err@extrafootnotemark{\immediate\write\sixt@@n
     {--> \noexpand\footnotemark command
          has no corresponding \noexpand\footnotetext}}

\def\err@extrafootnotetext{%
     \errhelp{You have given a \noexpand\footnotetext command without first
          specifying^^Ja \noexpand\footnotemark.^^J}%
     \errmessage{--> \noexpand\footnotetext command has no corresponding
          \noexpand\footnotemark}}

\def\err@labelredef#1{\immediate\write\sixt@@n
     {--> Label "#1" redefined}}

\def\err@badlabelmatch#1{\immediate\write\sixt@@n
     {--> Definition of label "#1" doesn't match value in \jobname.lab}}

\def\err@needlabel#1{\immediate\write\sixt@@n
     {--> Label "#1" cited before its definition}}

\def\err@undefinedlabel#1{\immediate\write\sixt@@n
     {--> Label "#1" cited but never defined}}

\def\err@undefinedeqn#1{\immediate\write\sixt@@n
     {--> Equation "#1" not defined}}

\def\err@undefinedref#1{\immediate\write\sixt@@n
     {--> Reference "#1" not defined}}

\def\err@nostream#1{%
     \errhelp={You have tried to input a stream file that doesn't exist.^^J}%
     \errmessage{--> Stream file #1 not found}}

\message{jyTeX initialization}

\everyjob{\immediate\write16{--> jyTeX version \fmtversion}%
     \edef\@@jobname{\jobname}%
     \edef\jobname{\@@jobname}%
     \settime
     \openin0=\jobname.lab
     \ifeof0
     \else\closein0
          \immediate\write16{--> Getting labels from file \jobname.lab}%
          \input\jobname.lab
     \fi}


%
     \^^\{\splittopskip}%
     \^^\{\maxdepth}%
     \^^\{\skip\topins}%
     \^^\{\skip\footins}%
     \^^\{\headskip}%
     \^^\{\footskip}}

\def\scalingskipslist{%
     \^^\{\p@renwd}%
     \^^\{\delimitershortfall}%
     \^^\{\nulldelimiterspace}%
     \^^\{\scriptspace}%
     \^^\{\jot}%
     \^^\{\normalbaselineskip}%
     \^^\{\normallineskip}%
     \^^\{\normallineskiplimit}%
     \^^\{\baselineskip}%
     \^^\{\lineskip}%
     \^^\{\lineskiplimit}%
     \^^\{\bigskipamount}%
     \^^\{\medskipamount}%
     \^^\{\smallskipamount}%
     \^^\{\parskip}%
     \^^\{\parindent}%
     \^^\{\abovedisplayskip}%
     \^^\{\belowdisplayskip}%
     \^^\{\abovedisplayshortskip}%
     \^^\{\belowdisplayshortskip}%
     \^^\{\abovechapterskip}%
     \^^\{\belowchapterskip}%
     \^^\{\abovesectionskip}%
     \^^\{\belowsectionskip}%
     \^^\{\abovesubsectionskip}%
     \^^\{\belowsubsectionskip}}


\def\twoupsetup{
     \topmargin=.75in
     \leftmargin=.5in
     \vsize=6.9in
     \hsize=4.75in
     \fullhsize=10in
     \let\draft=\relax}


\chapterstyle{left}                              
\chapternumstyle{blank}                          
\def\chapterbreak{\newpage}                      
\abovechapterskip=0pt                            
\belowchapterskip=1.5\baselineskip               
     plus.38\baselineskip minus.38\baselineskip
\def\chapternumformat{\numstyle\chapternum.}     

\sectionstyle{left}                              
\sectionnumstyle{blank}                          
\def\sectionbreak{\vskip0pt plus4\baselineskip\penalty-100
     \vskip0pt plus-4\baselineskip}              
\abovesectionskip=1.5\baselineskip               
     plus.38\baselineskip minus.38\baselineskip
\belowsectionskip=\the\baselineskip              
     plus.25\baselineskip minus.25\baselineskip
\def\sectionnumformat{
     \ifblank\chapternumstyle\then\else\numstyle\chapternum.\fi
     \numstyle\sectionnum.}

\subsectionstyle{left}                           
\subsectionnumstyle{blank}                       
\def\subsectionbreak{\vskip0pt plus4\baselineskip\penalty-100
     \vskip0pt plus-4\baselineskip}              
\abovesubsectionskip=\the\baselineskip           
     plus.25\baselineskip minus.25\baselineskip
\belowsubsectionskip=.75\baselineskip            
     plus.19\baselineskip minus.19\baselineskip
\def\subsectionnumformat{
     \ifblank\chapternumstyle\then\else\numstyle\chapternum.\fi
     \ifblank\sectionnumstyle\then\else\numstyle\sectionnum.\fi
     \numstyle\subsectionnum.}


\def\undefinedlabelformat{$\bullet$}             


\equationnumstyle{arabic}                        
\subequationnumstyle{blank}                      
\figurenumstyle{arabic}                          
\subfigurenumstyle{blank}                        
\tablenumstyle{arabic}                           
\subtablenumstyle{blank}                         
\defnumstyle{arabic}                             
\subdefnumstyle{blank}                           
\thmnumstyle{arabic}                             
\subthmnumstyle{blank}                           
\lemnumstyle{arabic}                             
\sublemnumstyle{blank}                           

\eqnseriesstyle{alphabetic}                      
\figseriesstyle{alphabetic}                      
\tblseriesstyle{alphabetic}                      
\defseriesstyle{alphabetic}                      
\thmseriesstyle{alphabetic}                      
\lemseriesstyle{alphabetic}                      

\def\puteqnformat{\hbox{
     \ifblank\chapternumstyle\then\else\numstyle\chapternum.\fi
     \ifblank\sectionnumstyle\then\else\numstyle\sectionnum.\fi
     \ifblank\subsectionnumstyle\then\else\numstyle\subsectionnum.\fi
     \numstyle\equationnum
     \numstyle\subequationnum}}
\def\putfigformat{\hbox{
     \ifblank\chapternumstyle\then\else\numstyle\chapternum.\fi
     \ifblank\sectionnumstyle\then\else\numstyle\sectionnum.\fi
     \ifblank\subsectionnumstyle\then\else\numstyle\subsectionnum.\fi
     \numstyle\figurenum
     \numstyle\subfigurenum}}
\def\puttblformat{\hbox{
     \ifblank\chapternumstyle\then\else\numstyle\chapternum.\fi
     \ifblank\sectionnumstyle\then\else\numstyle\sectionnum.\fi
     \ifblank\subsectionnumstyle\then\else\numstyle\subsectionnum.\fi
     \numstyle\tablenum
     \numstyle\subtablenum}}
\def\putdefformat{\hbox{
     \ifblank\chapternumstyle\then\else\numstyle\chapternum.\fi
     \ifblank\sectionnumstyle\then\else\numstyle\sectionnum.\fi
     \ifblank\subsectionnumstyle\then\else\numstyle\subsectionnum.\fi
     \numstyle\defnum
     \numstyle\subdefnum}}
\def\putthmformat{\hbox{
     \ifblank\chapternumstyle\then\else\numstyle\chapternum.\fi
     \ifblank\sectionnumstyle\then\else\numstyle\sectionnum.\fi
     \ifblank\subsectionnumstyle\then\else\numstyle\subsectionnum.\fi
     \numstyle\thmnum
     \numstyle\subthmnum}}
\def\putlemformat{\hbox{
     \ifblank\chapternumstyle\then\else\numstyle\chapternum.\fi
     \ifblank\sectionnumstyle\then\else\numstyle\sectionnum.\fi
     \ifblank\subsectionnumstyle\then\else\numstyle\subsectionnum.\fi
     \numstyle\lemnum
     \numstyle\sublemnum}}


\referencestyle{sequential}                      
\referencenumstyle{arabic}                       
\def\putrefformat{\numstyle\referencenum}        
\def\referencenumformat{\numstyle\referencenum.} 
\def\putreferenceformat{
     \everypar={\hangindent=1em \hangafter=1 }%
     \def\\{\hfil\break\null\hskip-1em \ignorespaces}%
     \leftskip=\refnumindent\parindent=0pt \interlinepenalty=1000 }


\def\fmtversion{2.6M (June 1992)}


\def\ref#1{(\puteqn{#1})}
\def\label#1{\eqno\eqnlabel{#1}}
\font\bigboldfont=cmbx10 scaled \magstep2
\font\boldfont=cmbx10 scaled \magstep1
\def\displayhead#1{{\bigboldfont \leftline{#1}}
\vskip-10pt
\line{\hrulefill}}
\def\section#1{\ifblank\sectionnumstyle\then
          \else\newsectionnum=\next \fi
\displayhead{\ifblank\sectionnumstyle\then\else\sectionnumformat\ \fi#1}
     }
\def\subsection#1{\advance\subsectionnum by 1
  {\boldfont \leftline{\ifblank\sectionnumstyle\then\else\sectionnumformat\fi\number\subsectionnum\
        #1}} \vskip-10pt \line{\hrulefill}}
\def\appendix#1{\ifblank\sectionnumstyle\then
          \else\newsectionnum=\next \fi
\displayhead{Appendix
    \ifblank\sectionnumstyle\then\else\sectionnumformat\ \fi#1}
     }


\catcode`\@=12